\documentclass[superscriptaddress,
reprint,
 amsmath,amssymb,
prb,
]{revtex4-1}
\usepackage{graphicx}
\usepackage{dcolumn}
\usepackage{bm}
\usepackage{subfig}

\usepackage[colorlinks=true,linkcolor=blue,citecolor=blue,urlcolor=blue]{hyperref}
\usepackage{braket}
\DeclareMathOperator{\T}{T}
\DeclareMathOperator{\tr}{tr}

\DeclareMathOperator{\HF}{HF}
\DeclareMathOperator{\QP}{QP}
\newcommand{\MP}{\mathrm{MP}}
\DeclareMathOperator{\occ}{occ}

\newcommand{\BW}{\mathrm{BW}}
\newcommand{\QPMP}{\mathrm{QPMP}}
\usepackage{bbm}
\newcommand{\DSRG}{\mathrm{DSRG}}
\newcommand{\phys}{\mathrm{phys}}
\newcommand{\PT}{\mathrm{PT}}

\begin{document}


\title{A regularized second-order correlation method from Green's function theory}

\author{Christopher J. N. Coveney}
\affiliation{ Department of Physics, University of Oxford, Parks Road, OX1 3PJ, United Kingdom}
\author{David P. Tew}%
\email{david.tew@chem.ox.ac.uk}
\affiliation{Physical and Theoretical Chemistry Laboratory, University of Oxford, South Parks Road, OX1 3QZ, United Kingdom\\
\emph{Email: \url{david.tew@chem.ox.ac.uk}}}

\date{\today}

\begin{abstract}
We present a scalable single-particle framework to treat electronic correlation in molecules and materials motivated by Green's function theory. We derive a size-extensive Brillouin-Wigner perturbation theory from the single-particle Green's function by introducing the Goldstone self-energy. This new ground state correlation energy, referred to as Quasi-Particle MP2 theory (QPMP2), avoids the characteristic divergences present in both second-order Møller-Plesset perturbation theory and Coupled Cluster Singles and Doubles within the strongly correlated regime. We show that the exact ground state energy and properties of the Hubbard dimer are reproduced by QPMP2 and demonstrate the advantages of the approach for the six-, eight- and ten-site Hubbard models where the metal-to-insulator transition is qualitatively reproduced, contrasting with the complete failure of traditional methods. We apply this formalism to characteristic strongly correlated molecular systems and show that QPMP2 provides an efficient, size-consistent regularization of MP2.
\end{abstract}


\maketitle


\section{\label{sec:intro}\protect Introduction}
The characteristic divergences of second-order Møller-Plesset perturbation theory (MP2) and CCSD for typical condensed-matter and molecular systems represent a significant unsolved problem in electronic structure theory.\cite{scuseria2008ground,shepherd2013many,bochevarov2005hybrid} A scalable, quantitatively accurate and non-divergent electronic structure theory for strong correlation remains elusive. Whilst CCSD is a successful many-body theory of electronic correlation, the origin of its divergences stems from its dependence on the Hartree-Fock (HF) reference determinant.\cite{kowalski2000renormalized}

The divergences of MP2 theory for insulators arise predominantly due to the denominator of the correlation energy. At internuclear separations where the HF single-particle energies are similar and the coulomb matrix elements remain finite, the theory will clearly diverge. Historically, the issue of this divergence has been handled by introducing various regularization schemes. The $\kappa$-regularization scheme of Lee and Head-Gordon has attracted a lot of interest
\begin{gather}
    \begin{split}\label{eq:kappa}
        E^{\kappa - \MP2}_{c}(\kappa) &= -\frac{1}{4}\sum_{ia,jb} \frac{|\braket{ij||ab}|^2}{\Delta^{ab}_{ij}}\left(1-e^{-\kappa\Delta^{ab}_{ij}}\right)^2 \\
        \Delta^{ab}_{ij} &= \epsilon_a + \epsilon_b - \epsilon_i -\epsilon_j \ ,
    \end{split}
\end{gather}
where $\{\epsilon_p\}$ are the Fock energies, $i$ and $a$ index occupied and unoccupied Fock orbital states, $\braket{ij||ab}$ is the anti-symmetrized Coulomb integral and $\kappa$ is empirically determined though benchmarking datasets.\cite{loipersberger2021exploring} This regularization scheme is useful in curing the divergent behaviour of MP2 when the divergences are caused by the vanishing denominator $\Delta^{ab}_{ij}$. However, the $\kappa$-regularization scheme diverges for the Hubbard model and therefore cannot account for the qualitative metal-to-insulator transition observed as $\frac{U}{t}$ increases (Section \ref{hubbard}).\cite{keller2022regularized}

Another approach to regularization takes the form of Brillouin-Wigner perturbation theory (BWPT) which is exact for two-level systems and avoids the divergence of the MP2 denominator.\cite{papp2007many} However, BWPT is not size-extensive and gives multiple solutions which are non-desirable characteristics.  

The Driven Similarity Renormalization Group (DSGR) method of Evangelista and co-workers can also be viewed as another size-consistent regularization of second-order perturbation theory.\cite{evangelista2014driven,li2015multireference,li2017driven,li2019multireference} At second-order in perturbation theory of the similarity transformed Hamiltonian obtained from the DSRG transformation, the correlation energy is given by\cite{evangelista2014driven}
\begin{equation}\label{eq:DSRG2}
E^{\DSRG-\PT2}_{c}(s) = -\frac{1}{4}\sum_{ia,jb} \frac{|\braket{ij||ab}|^2}{\Delta^{ab}_{ij}}\left[1-e^{-2s(\Delta^{ab}_{ij})^2}\right]  \ . 
\end{equation}
where $s \in [0,\infty)$ is the flow parameter which is related to an energy cutoff, $\Lambda = s^{-\frac{1}{2}}$. The flow parameter essentially decouples states of the Hamiltonian which are separated by an energy difference, $\Delta^{ab}_{ij}$ larger than the energy cutoff, thereby alleviating the intruder state problem.\cite{camacho2009intruder, monino2022unphysical, evangelisti1987qualitative,battaglia2022regularized} The advantages of the DSRG method stem from the fact that it can be formulated as a modified unitary coupled cluster theory which allows for iterative solution of the flow equations thereby including interactions beyond second-order perturbation theory.\cite{evangelista2014driven} Additionally, the method can also be formulated based on a generalized reference wavefunction referred to as DSRG-MRPT.\cite{li2019multireference} However, whilst the regularized second-order perturbation theory in expression \ref{eq:DSRG2} (DSRG-PT2) cures the divergent behaviour of MP2 theory caused by the vanishing denominator, it also diverges akin to the $\kappa$ regularization scheme for the metal-to-insulator transition present in the Hubbard model as $\frac{U}{t}$ increases. 

Motivated by these issues, we introduce a size-consistent and natural second-order regularization scheme based on the single-particle Green's function. This implementation is rooted in the second-order Goldstone self-energy which we refer to as quasi-particle MP2 theory (QPMP2). Importantly, this provides us with a first principles regularization which recovers the performance of MP2 in the weakly correlated regime, whilst extending its domain of applicability to the strongly correlated regime. Our results show that QPMP2 recovers the metal-to-insulator phase behaviour of various Hubbard models and extends the domain of MP2 theory for strongly correlated molecular systems. However, we find that for Green's functions methods in general to be successful in quantum chemistry requires higher-order interaction terms to be taken into account. 

The Green's function formalism represents an effective Hamiltonian theory resulting from a field theoretic approach to the electronic structure problem.\cite{Quantum,mahan2000many,stefanucci2013nonequilibrium} By working directly with the second-quantized electron field operators themselves, the Green's function appears naturally when computing expectation values of many observables. The advent of quantum field theory and quantum electrodynamics introduced many sophisticated methods for evaluating the single-particle Green's function based on time-dependent perturbation theory.\cite{dyson1949radiation,dyson1949s,martin1959theory,gell1951bound,feynman1949space} By folding in the large numbers of degrees of freedom into the dynamical self-energy, a single-particle potential which contains all the many-body interactions present in the system, the many-body problem is expressed entirely in terms of a `one-particle' theory. Of course, there are many other effective Hamiltonian approaches operating in both the single-particle and many-body formalism, which have been explored extensively in the literature.\cite{crawford2007introduction,bartlett2007coupled,offermann1976degenerate,stanton1993equation,andersson1990second,kozlowski1994considerations} In this work, we focus on the second-order Green's function perturbation theory approach, where the dynamical self-energy is approximated to second-order in the interaction, as a means of regularizing the MP2 correlation energy.

The second-order, self-consistent Green's function perturbation theory (GF2) has also been shown to cure the divergences of MP2 in a size-consistent, single-particle framework and has been extremely successful in providing a self-consistent treatment of strongly correlated systems. However, this approach is implemented within a temperature dependent Matsubara frequency formulation and is a much more involved numerical implementation deriving from statistical mechanics.\cite{dzyaloshinski1975methods,kananenka2016efficient,neuhauser2017stochastic,dahlen2005self,nguyen2016rigorous} The ground state correlation energy evaluated within the GF2 implementation of Zgid and co-workers is explicitly temperature dependent and obtained by evaluation of the Matsubara frequency summation of a modified Galitskii-Migdal (GM) formula dependent on both the real and imaginary parts of the Green's function and self-energy.\cite{phillips2014communication} The implementation is performed entirely within the atomic orbital basis and by building the self-energy on the imaginary time grid which scales with the number of imaginary time grid points used. The subsequent Fourier transformations and self-consistency requirements result in an increase in the formal scaling of the GF2 implementation as $\mathcal{O}(n_{\tau}n_{at}^5)$ per iteration, where $n_\tau$ is the number of imaginary time grid points and $n_{at}$ is the number of atomic orbitals. 

The auxiliary GF2 method provides an alternative approach to the renormalized second-order self-energy by exploiting the fact that the effects of the self-energy can be expressed by coupling the system to a set of fictitious external degrees of freedom spanned by the so-called `auxiliary' states.\cite{backhouse2020efficient} This approach reformulates the Dyson equation in the language of wavefunctions and avoids the use of frequency grids and numerical Fourier transforms.\cite{backhouse2020wave,backhouse2020efficient} The ground state correlation energy is self-consistently evaluated using a modified Galitskii-Migdal formula at zero temperature to generate a renormalized MP2-like correlation energy derived from the auxiliary degrees of freedom. The regularized MP2 correlation energy introduced in auxiliary-GF2 is obtained by identifying that directly applying the GM formula gives twice the MP2 energy for the non-interacting Green's function.\cite{backhouse2020efficient} However, this regularized correlation energy is not directly obtained from the self-consistent Green's function formalism, but is rather obtained by halving the two-body correlation energy obtained from the GM formula using the `interacting' Green's function containing the auxiliaries. Purely using Green's function auxiliary compression means that each iteration, until self-consistency is obtained, would scale as $\mathcal{O}(N_{\phys}^9)$ where $N_{\phys}$ is the number of spin-orbitals in the physical subspace. With a two-step auxiliary compression scheme (which introduces a small error in the true correlation energy obtained from the full procedure) this scaling can be formally reduced to 
$\mathcal{O}(N_\phys^5)$ \emph{i.e.} the scaling of regular MP2 theory at each iteration until self-consistency is obtained.

Our contribution connects BWPT and MP2 regularization methods with the second-order frequency dependent self-energy in a clear and simple way. By employing the quasi-particle approximation for the Green's function and introducing the forward time second-order Goldstone self-energy, we are able to show directly how the regularized MP2 correlation energy naturally appears. This sheds light on how second-order approximations for the self-energy act to regularize the MP2 correlation energy and allows us to investigate the effects of the quasi-particle approximation on ground state correlation energies. Our approach considers the effects of the dynamical second-order self-energy only on the occupied states by noting the similarity of the ground state density obtained from the quasi-particle Green's function with that of Kohn-Sham Density Functional Theory (KS-DFT). Our QPMP2 method takes the quasi-particle solution from the second-order Goldstone self-energy and uses these renormalized single-particle energies to calculate the electronic correlation from the associated Green's function. From this approach, we are able to cure the divergences present in MP2 whilst remaining within a scaling of $\mathcal{O}(N^5)$, where $N$ is the number of spin-orbitals. Our regularization scheme is particularly adapt for Hubbard model systems, where both the $\kappa$ and DSRG-PT2 regularization schemes as well as CCSD diverge. This is a dramatic result which demonstrates the regularization of second-order perturbation theory afforded by QPMP2 across the entire correlation regime. Our results for molecular systems show that this simple method provides similar performance to MP2 within the weakly correlated regime, but extends its domain of applicability when the MP2 solution diverges in the strongly correlated regime. To the best of our knowledge, this approach to regularization has not yet been explored, but is closely related to GF2 and other second-order regularization schemes.\cite{holleboom1990comparison,keller2022regularized}

The structure of this paper is as follows. Section \,\ref{green} provides background on the single-particle Green's function and its relationship to the ground state energy. Section\, \ref{qp} outlines the structure of quasi-particle Green's function theory and the solutions of the quasi-particle equations. Section \,\ref{goldstone} provides the definition of the Goldstone self-energy which contains the interactions encoded by QPMP2 theory. Section \,\ref{BW} outlines how our approach can be viewed as a size-extensive Brillouin-Wigner perturbation theory. In Sections \ref{hubbard} and \ref{molecules} we apply QPMP2 to Hubbard models and molecular systems, comparing with other electronic structure theories. Finally in Section \ref{conclusion}, we summarise the results of our method and outlook for future scalable electronic structure theories.

\section{Single-particle Green's Function and Ground State Energy}\label{green}

The second-quantized, non-relativistic electronic structure Hamiltonian in atomic units and within the Born-Oppenheimer approximation is given by\cite{Quantum}
\begin{gather}\label{eq:1.2}
\begin{split}
     H&=\int d\mathbf{x}\ \psi^{\dag}(\mathbf{x})h(\mathbf{r})\psi(\mathbf{x})\\
     &+\frac{1}{2}\int d\mathbf{x}d\mathbf{x}'\ \psi^{\dag}(\mathbf{x})\psi^{\dag}(\mathbf{x}')v(\mathbf{r},\mathbf{r}')\psi(\mathbf{x}')\psi(\mathbf{x})\\
     h(\mathbf{r})&= -\frac{1}{2}\nabla^{2} + v_{ne}(\mathbf{r})\\
    v_{ne}(\mathbf{r})&=-\sum_{I}\frac{Z_{I}}{|\mathbf{r}-\mathbf{R}_{I}|}\\
    v(\mathbf{r},\mathbf{r}') &=\frac{1}{|\mathbf{r}-\mathbf{r}'|}
\end{split}
\end{gather}
where $\mathbf{x}_1 \equiv (\mathbf{r}_1,\sigma_1)$ is the composite spin-space coordinate and $\psi^{\dag}(\mathbf{x})/\psi(\mathbf{x})$ are the field operators that create/annihilate electrons at the composite spin-space coordinate $\mathbf{x}$. The sets $\{\mathbf{r}_{i}\}$ and $\{\mathbf{R}_{I}\}$ refer to the electron and nuclear coordinates respectively. $\{Z_{I}\}$ is the set of nuclear charges associated with the nuclei. The single particle Green's function is defined by the following time-ordered expectation value 
\begin{equation}
    iG(\mathbf{x}_1t_1,\mathbf{x}_2t_2) = \braket{N,0|\T\left[\psi(\mathbf{x}_1t_1)\psi^{\dag}(\mathbf{x}_2t_2)\right]|N,0}
\end{equation}
where $\ket{N,0}$ is the normalized,
exact $N$-electron ground state and $\T$ is the time-ordering operator which places operators with the larger time 
argument to the left. The field operators are defined within the Heisenberg picture. From the Green's 
function, it is possible to find the ground state expectation value of any single-particle operator
\begin{equation}
    \braket{O}_0 = -i\int d\mathbf{x}_1 d\mathbf{x}_2\ o(\mathbf{x}_1,\mathbf{x}_2)G(\mathbf{x}_2t_2,\mathbf{x}_1t_2^+)
\end{equation}
where $t^+_2 = \lim_{\eta\to0^+}t_2+\eta$ where $\eta$ tends to zero from above. From the equation of motion for the Green's function it is also possible to extract the total ground state energy
\begin{gather}
    \begin{split}\label{eq:energy}
        E^{N}_0 &= \frac{1}{2i} \int d\mathbf{x}_1 \lim_{t_2\to t_1^+}\left[i\frac{\partial}{\partial t_1}+h(\mathbf{r}_1)\right]G(\mathbf{x}_1t_1,\mathbf{x}_1t_2)\\
        &= \lim_{\eta\to 0^+}\frac{1}{2} \int d\mathbf{x}_1 \int^{\infty}_{-\infty} \frac{d\omega}{2\pi i} \left[\omega+h(\mathbf{r}_1)\right]G(\mathbf{x}_1,\mathbf{x}_1;\omega)e^{i\omega\eta}\\
        &= \lim_{\eta\to 0^+}\frac{1}{2}\int^{\infty}_{-\infty} \frac{d\omega}{2\pi i} e^{i\omega\eta}\tr\{\left(\omega+\mathbf{h}\right)\mathbf{G}(\omega)\}.
    \end{split}
\end{gather}
In general, the Green's function can be constructed from its Lehmann representation as
\begin{gather}
    \begin{split}\label{eq:green}
        G(\mathbf{x}_1,\mathbf{x}_2;\omega) &= \sum_{a} \frac{\psi_{a}(\mathbf{x}_1)\psi^{*}_{a}(\mathbf{x}_2)}{\omega-\varepsilon_{a}+i\eta}
        +\sum_{j} \frac{\psi_{j}(\mathbf{x}_1)\psi^{*}_{j}(\mathbf{x}_2)}{\omega-\varepsilon_{j}-i\eta}
    \end{split}
\end{gather}
where $\psi_j(\mathbf{x})=\braket{N-1,j|\psi(\mathbf{x})|N,0}$ and $\psi_a(\mathbf{x})= \braket{N,0|\psi(\mathbf{x})|N+1,a}$ are the `single-particle' Dyson orbital states. $\varepsilon_a = E^{N+1}_a - E^N_0$ and $\varepsilon_j = E^{N}_0-E^{N-1}_j$ are the exact electron addition and removal energies respectively. From the Lehmann representation, the Green's function reduces to the particle density via 
\begin{equation}\label{eq:density}
    \rho(\mathbf{r}) = \lim_{\eta\to0^+}\int d\sigma \int^{\infty}_{-\infty} \frac{d\omega}{2\pi i} e^{i\omega\eta} G(\mathbf{x},\mathbf{x};\omega) = \sum_{j} \psi_{j}(\mathbf{r})\psi^{*}_{j}(\mathbf{r}) \ .
\end{equation}
Importantly, this relation is non-invertible and thus it is impossible to obtain the Green's function from the density. It is clear from above that the ground state energy of a system of $N$-interacting electrons can be expressed in terms of the Dyson orbitals as\cite{galitskii1958application}
\begin{equation}
    E^N_0 = \frac{1}{2}\sum_{j} \Big[\braket{\psi_j|h|\psi_j} + \varepsilon_j\braket{\psi_j|\psi_j}\Big]
\end{equation}
In general, the non-orthogonal and overcomplete set of Dyson-orbitals are normalized by their pole strength or probability factors, $P_i$ \cite{ortiz2020dyson} 
\begin{gather}
    \begin{split}
        \int d\mathbf{x}\ &|\psi_i(\mathbf{x})|^2 = P_i \\
        \sum_{i} &P_i = N \\
        P_i &= \left(1-\frac{\partial\Sigma_{ii}(\omega)}{\partial\omega}\Big|_{\varepsilon_i}\right)^{-1} .
    \end{split}
\end{gather}
The probability factors are constrained between zero and one and the summation requires that the minimum number of ionization poles any Green's function possesses must be equal to the number of occupied spin-orbitals. In order to capture all the many-body correlation effects within a single-particle framework, it is necessary for the dimension of the Dyson spin-orbital basis to be larger than that of the spin-orbital one due to the one-to-one mapping between ionization processes and Dyson orbitals. The probability factors are the residues of the poles of the single-particle Green's function. In the HF approximation, these residues are all normalized to one for every pole of the Green's function and the number of ionization poles is equal to the total number of electrons. Under conditions where there is a dominant solution of the Dyson equation, one with the largest residue, the Green's function can be approximated by simply just taking this pole. In this case, we refer to any such Green's function as a quasi-particle Green's function. From equation (\ref{eq:density}), the ground state electronic density is a functional only of the `occupied' Dyson orbitals $\{\psi_i\}$, which is analogous to the expression for the density in KS-DFT. When working with the quasi-particle Green's function, the expression for the density is exactly that obtained from KS-DFT (a sum over $N$ orthonormal  spin-orbitals).  

\section{Quasi-particle Green's Function Theory}\label{qp}
 The equation of motion for the Green's function can be transformed into the Dyson equation. Written in the spin-orbital basis and frequency domain this takes the form of 
\begin{equation}\label{dyson}
    \mathbf{G}^{-1}(\omega) = \mathbf{G}^{-1}_0(\omega)-\mathbf{\Sigma}(\omega) \ ,
\end{equation}
where $\mathbf{\Sigma}(\omega)$ is the dynamical self-energy containing all the many-body interactions experienced by a single-particle and $\mathbf{G}_0(\omega)$ is a reference Green's function which we take here to be the HF propagator. This equation of motion can be re-cast as the quasi-particle equation 
\begin{equation}\label{eqn:qp-eqn}
    \left[f+\Sigma(\varepsilon_{p})\right]\ket{\psi_{p}} = \varepsilon{_{p}}\ket{\psi_p} 
\end{equation}
where $f$ is the Fock operator. 
The poles of the interacting Green's function are given by the excitation energies $\varepsilon_{p}$ which are the energies at which the inverse single-particle Green's function vanishes. As a result, the solutions of equation \ref{eqn:qp-eqn} yields the exact poles and residues of the single-particle Green's function, meaning that we need only find these solutions to completely specify the single-particle Green's function (equation \ref{eq:green}). To find the quasi-particle solutions ($\{\epsilon^{\QP}_p\}$), which are those of the largest residue, we can express the quasi-particle equation in terms of a perturbation theory with respect to the Fock orbitals
\begin{equation}\label{eq:qp-pert}
    f+\lambda\Sigma(\epsilon^{\QP}_p)\ket{\psi_p} = \epsilon^{\QP}_{p}\ket{\psi_p} \ .
\end{equation}
For each Fock wavefunction we define two projectors
\begin{equation}
    P_p = \ket{\phi_p}\bra{\phi_p} \ ;\ Q_{p} = \sum_{p'\neq p}\ket{\phi_{p'}}\bra{\phi_{p'}}
\end{equation}
where $f\ket{\phi_p} = \epsilon_p\ket{\phi_p}$. Resolving the identity onto the quasi-particle wavefunction, we have
\begin{equation}
    \ket{\psi_p} = \braket{\phi_p|\psi_p}\ket{\phi_p} + Q_p\ket{\psi_p} 
\end{equation}
We can normalize the Fock single-particle state with the quasi-particle state by letting $\braket{\phi_p|\psi_p}=1$ as $\ket{\phi_p}$ and $\ket{\psi_p}$ correspond to each other. Treating the self-energy as a perturbation to the Fock operator, we apply $Q_p$ to both sides of equation \ref{eq:qp-pert} and commuting it past $f$ gives
\begin{equation}\label{eq:pert}
    \lambda\frac{Q_p}{\epsilon^{\QP}_p-f}\Sigma(\epsilon^{\QP}_p)\ket{\psi_p} = Q_p\ket{\psi_p} \ .
\end{equation}
Therefore, from equation \ref{eq:pert}, we have the Dyson equation for the quasi-particle wavefunction
\begin{gather}
\begin{split}
        \ket{\psi_p} &= \ket{\phi_p} + \lambda\frac{Q_p}{\epsilon^{\QP}_p-f}\Sigma(\epsilon^{\QP}_p)\ket{\psi_p}\\
        &= \ket{\phi_p} + \lambda\frac{Q_p}{\epsilon^{\QP}_p-f}\Sigma(\varepsilon_p)\ket{\phi_p} + \\
        &\lambda^2\frac{Q_p}{\epsilon^{\QP}_p-f}\Sigma(\epsilon^{\QP}_p)\frac{Q_p}{\epsilon^{\QP}_p-f}\Sigma(\epsilon^{\QP}_p)\ket{\phi_p} + \mathcal{O}(\lambda^3)
\end{split}
\end{gather}
The resulting equation for the quasi-particle energy is given by 
\begin{equation}
    \bra{\phi_p}\left[f+\lambda\Sigma(\epsilon^{\QP}_p)\right]\ket{\psi_p} = \bra{\phi_p}\epsilon^{\QP}_{p}\ket{\psi_p}
\end{equation}
which, using the normalization above gives
\begin{gather}
\begin{split}\label{eq:series}
        \epsilon^{\QP}_p &= \epsilon_p + \lambda\braket{\phi_p|\Sigma(\epsilon^{\QP}_p)|\phi_p} + \\
        &\lambda^2\sum_{p'\neq p} \frac{\braket{\phi_{p}|\Sigma(\epsilon^{\QP}_{p})|\phi_{p'}}\braket{\phi_{p'}|\Sigma(\epsilon^{\QP}_p)|\phi_p}}{\epsilon^{\QP}_p-\epsilon_{p'}} + ...
\end{split}
\end{gather}
The self-energy can also be expanded perturbatively in the order of interaction (Feynman diagram series) as follows 
\begin{equation}
    \Sigma(\epsilon^{\QP}_p) = \sum_{n = 1}^{\infty} \mu^n \Sigma^{(n)}(\epsilon^{\QP}_p)
\end{equation}
where $\Sigma^{(1)} = 0$ as the zeroth-order Green's function corresponds to the Fock operator. Inserting this expression into \ref{eq:series} gives  
\begin{gather}
    \begin{split}\label{eq:dyson_series}
        &\epsilon^{\QP}_p = \epsilon_p + \lambda\sum_{n=1}^{\infty}\mu^{n}\braket{\phi_p|\Sigma^{(n)}(\epsilon^{\QP}_p)|\phi_p} + \\
        &\lambda^2\sum_{n,m=1}^{\infty}\mu^{n+m}\sum_{p'\neq p} \frac{\braket{\phi_{p}|\Sigma^{(n)}(\epsilon^{\QP}_{p})|\phi_{p'}}\braket{\phi_{p'}|\Sigma^{(m)}(\epsilon^{\QP}_p)|\phi_p}}{\epsilon^{\QP}_p-\epsilon_{p'}} \\
        & + ...
    \end{split}
\end{gather}
Equation \ref{eq:dyson_series} is simply a restatement of the Dyson geometric series summation of equation \ref{dyson} as a function of the self-energy perturbation. As the self-energy contains only connected diagrams, no reducible diagram contributions appear. Expanding the self-energy to second-order in the interaction we have 
\begin{gather}
    \begin{split}
        \epsilon^{\QP}_p = \epsilon_p + \lambda\mu^{2}\braket{\phi_p|\Sigma^{(2)}(\epsilon^{\QP}_p)|\phi_p} + \mathcal{O}\left(\lambda^2\mu^{4}\right) \ ,
    \end{split}
\end{gather}
which, by setting $\lambda=\mu=1$ gives 
\begin{equation}
    \epsilon^{\QP}_p = \epsilon_p +\braket{\phi_p|\Sigma^{(2)}(\epsilon^{\QP}_p)|\phi_p} \ .
\end{equation}
We have converted the quasi-particle equation into a Brillouin-Wigner perturbation theory for the corrected quasi-particle energies from the Fock reference states. The second-order quasi-particle equations are written as 
\begin{equation}\label{eq:BW}
    \epsilon^{\QP}_{p}\delta_{pq} = \epsilon_{p}\delta_{pq} + \Sigma^{(2)}_{pp}(\epsilon^{\QP}_{p})\delta_{pq} \ 
\end{equation}
which is equivalent to the diagonal approximation of the Dyson equation in equation \ref{dyson}. The diagonal approximation is regularly used in $GW$ band structure calculations to compute the corrected addition and removal energies from a DFT calculation.\cite{hybertsen1986electron,rohlfing2000electron} In this case, the self-energy is given by the $GW$ approximation with the polarization calculated within the random phase approximation (RPA).\cite{hedin1965new,aryasetiawan1998gw,reining2018gw} However, usually equation \ref{eq:BW} is solved in a linearized form by Taylor expansion of the $GW$ self-energy about the KS-DFT energy which also introduces renormalization factors.\cite{van2013gw,kaplan2016quasi} Additionally, these calculations are not concerned with the implications for the ground state correlation energy.\cite{van2015gw} The solutions of equation \ref{eq:BW} within the $GW$ approximation have been explored extensively in the form of ev$GW$ and have been shown to be comparable to those of the fully self-consistent qs$GW$.\cite{kaplan2016quasi} Importantly, the self-energy appearing in equation \ref{eq:BW} is expanded as the Feynman-Dyson self-energy functional $\Sigma[G,v]$, \textbf{not} in terms of the renormalized self-energy functional appearing in Hedin's equations ($\Sigma[G,W]$).  

Our choice to normalize the HF single-particle state with the quasi-particle state means we have made a quasi-particle transformation where there is a one-to-one correspondence between the HF orbitals and the quasi-particle orbitals. Taylor expanding the self-energy about the HF energy gives 
\begin{gather}
    \begin{split}
        \Sigma^{(2)}_{pp}(\epsilon^{\QP}_p) \approx \Sigma^{(2)}_{pp}(\epsilon_p) + \frac{\partial\Sigma^{(2)}_{pp}(\omega)}{\partial\omega}\Big|_{\epsilon_p}\left(\epsilon^{\QP}_p-\epsilon_p\right) + ...
    \end{split}
\end{gather}
and taking the zeroth order term gives 
\begin{equation}\label{eq:MP}
    \epsilon^{\QP}_{p}\delta_{pq} = \epsilon_{p}\delta_{pq} + \Sigma^{(2)}_{pp}(\epsilon_{p})\delta_{pq} \ .
\end{equation}
We shall see that this expression simply gives the MP2 correlation energy with the Goldstone self-energy (Section \ref{goldstone}). However, working with expression (\ref{eq:BW}) will give us a well-defined, simple and size consistent expression for the ground state correlation energy: QPMP2. In practise, solving for the quasi-particle energy in equation \ref{eq:BW} simply amounts to finding the largest root.

By taking the quasi-particle solutions only, the Green's function represents a canonical transformation. This means that we need only find the root with the largest residue which can easily be done via an iterative solver. This substantially simplifies the issue of having to calculate all the solutions of this non-linear equation and significantly improves the scaling. Inclusion of all the satellite solutions of the quasi-particle equation corresponding to less significant many-body interactions and their residues constitutes a significant increase in computational effort. We show from our Hubbard model results that inclusion of these satellites, required to construct the full Green's function, makes little to no impact on the ground state correlation energy within the second-order approximation.

\section{The Goldstone Exchange-Correlation Self-Energy}\label{goldstone}

In this section we define the second-order self-energy used to solve the quasi-particle equation (\ref{eq:BW}). Our choice of self-energy provides a straightforward connection to MP2 regularization methods and size-consistent Brillouin-Wigner perturbation theory. This leads us to the definition of the second-order self-energy in equation \ref{eq:SE} which enters the QPMP2 correlation energy derived in Section \ref{BW}.

We can generate a perturbation expansion for the self-energy of occupied states by expanding the exchange-correlation energy from the infinite Goldstone diagram series. This is analogous to the Luttinger-Ward functional introduced in the temperature dependent formalism.\cite{luttinger1961analytic,potthoff2014making} The Goldstone diagrams are defined with a pre-specified time direction with arrows moving upwards corresponding to virtual states (particles), whilst those moving downwards are occupied states (holes).\cite{Quantum} We define the self-energy as the functional derivative of the Goldstone correlation energy with respect to the hole Green's function. The Goldstone self-energy is identical to the forward time Feynman-Dyson self-energy up to third order.\cite{hirata2017one}

\begin{figure}[ht]
    \includegraphics[width=80mm, height = 35mm]{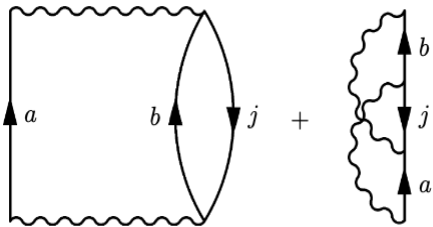} 
    \caption[Ground State Energy of Six-Site Model]{Second-order Goldstone self-energy diagrams. Time is defined upwards, particle states are labelled $a, b$ and hole states are labelled $j$.}
    \label{fig:gold}
\end{figure}

The self-energy of the occupied states simply corresponds to cutting the downward arrows of the Goldstone diagrams. This definition means that we can regain the Møller-Plesset perturbation theory expansion by simply evaluating the trace of the Goldstone self-energy at the single-particle HF energies.\cite{hirata2017one} At second-order, the Goldstone self-energy is given by cutting the downward arrows of the direct and exchange diagrams to yield the expression (Figure \ref{fig:gold})
\begin{equation}\label{eq:SE}
    \Sigma^{(2)}_{ii}(\omega) = \frac{1}{2} \sum_{abj}\frac{|\braket{ij||ab}|^2}{\omega+\epsilon_{j}-\epsilon_{a}-\epsilon_{b}} \ .
\end{equation}
This also ensures a real-valued self-energy with no lifetime effects. From this expression, we directly see that the self-energy is a frequency dependent single-particle potential containing many-body interactions present in the system. We regard this definition of the self-energy as being the most useful in the context of ground state electronic correlation. In Feynman-Dyson perturbation theory, the self-energy is a complex time-ordered quantity with real and imaginary parts.\cite{hirata2015general} This is the quantity that is approximated in the GF2 and auxillary GF2 methods. However, the presence of a complex self-energy no longer guarantees real energy solutions of the resulting quasi-particle equations. The real part of these solutions is usually interpreted as the renormalized single-particle energies whilst the imaginary part is interpreted as an exponential lifetime of the state.\cite{van2013gw,reining2018gw} However, using the Goldstone self-energy, we have a zero imaginary part - precisely the conditions under which Brillouin-Wigner perturbation theory is valid.\cite{mahan2000many} 

\section{Quasi-Particle MP2 as a Size-extensive Brillouin-Wigner Perturbation Theory}\label{BW}
In this section, we derive the QPMP2 correlation energy from the quasi-particle Green's function and outline how QPMP2 can be interpreted as a size-consistent formulation of second-order BWPT. The central result of this section is the QPMP2 correlation energy derived in equation \ref{qpmp2}.

In an orthornormal spin-orbital basis, the exact Green's function is defined as 
\begin{gather}
    \begin{split}
        G_{pq}(\omega) &= \sum_{i}\frac{C^i_{p}{C^{i}_q}^*}{\omega-\varepsilon_i-i\eta}+\sum_a \frac{D^a_{p}{D^{a}_q}^*}{\omega-\varepsilon_a+i\eta},
    \end{split}
\end{gather}
where $C^i_p = \braket{\phi_p|\psi_i}$ and $D^i_p = \braket{\phi_p|\psi_a}$ are the overlap between the Dyson orbitals and the occupied/virtual HF states and $\varepsilon_{i/a}$ are the exact electron removal and addition energies defined in Section \ref{green}. The ground state energy is given by the contour integral in equation \ref{eq:energy} which evaluates to
\begin{equation}
    E^N_0 = \frac{1}{2} \left(\sum_i \varepsilon_i \sum_p C^i_{p}{C^{i}_p}^* + \sum_i \sum_{pq} h_{pq}C^i_{q}{C^{i}_p}^* \right) . 
\end{equation}
The residues/probability factors are simply given by
\begin{equation}
    P_i = \sum_p C^i_{p}{C^{i}_p}^*  
\end{equation}
By choosing the quasi-particle normalization condition, we have that 
\begin{equation}
    \sum_{i} C^i_{q}{C^{i}_p}^* = \delta_{pq}
\end{equation}
as the set of quasi-particle states now constitute a complete set and the transformation is unitary. These states are obtained as the single largest solution of the quasi-particle equation as outlined in Section \ref{qp}. Therefore, the quasi-particle GF energy is given by 
\begin{gather}
    \begin{split}
        E^{N}_0 = \frac{1}{2}\left( \sum^{\occ}_{i} \epsilon^{\QP}_i + \sum^{\occ}_{i
    } h_{ii}\right)
    \end{split}
\end{gather}
Using equation \ref{eq:BW} as a very good first approximation, we see that the quasi-particle energy is given by 
\begin{equation}\label{eq:qp_eqn}
    \epsilon^{\QP}_i = \epsilon_i + \Sigma^{(2)}_{ii}(\epsilon^{\QP}_i)
\end{equation}
Therefore, the quasi-particle GF energy reduces to 
\begin{equation}
    E^N_0 = E^{\HF}_0 + \frac{1}{2} \sum^{\occ}_i \Sigma^{(2)}_{ii}(\epsilon^{\QP}_i) 
\end{equation}
From Section \ref{goldstone}, the exchange-correlation self-energy can be systematically approximated according to the Goldstone diagram series (equation \ref{eq:SE}). The second-order expression of equation \ref{eq:SE} gives the single-shot quasi-particle MP2 correlation energy (QPMP2) as
\begin{gather}
    \begin{split}\label{qpmp2}
        E^{\QPMP2}_c = \frac{1}{4}\sum_{ijab} \frac{|\braket{ij||ab}|^2}{\epsilon^{\QP}_i+\epsilon_{j}-\epsilon_{a}-\epsilon_{b}} \ .
    \end{split}
\end{gather}
This expression is clearly size extensive and represents a regularization of the MP2 expansion based on a Green's function approach. Although not pursued here, this expression can be trivially extended to finite temperature by working in the Matsubara formalism which introduces the relevant Fermi-Dirac occupation factors. Expression \ref{qpmp2} is symmetric with respect to particle exchange as the summation runs over both occupied dummy indices $i,j$. It is cast in the canonical Hartree--Fock orbital basis and while an orbitally invariant formulation is possible, the resulting equations are much more involved.
The MP2 correlation energy corresponds to taking the quasi-particle energy from the Møller-Plesset perturbation expression in equation \ref{eq:MP}. Therefore for our definition of the self-energy, all higher order MPn correlation energies can also be found by evaluating the trace of the nth-order Goldstone self-energy at the HF energies. 

Furthermore, one may iterate the self-energy as to include the effects casued by the renormalized occupied states.\cite{mahan2000many,Quantum,stefanucci2013nonequilibrium} 
\begin{figure}[ht]
    \includegraphics[width= 60mm]{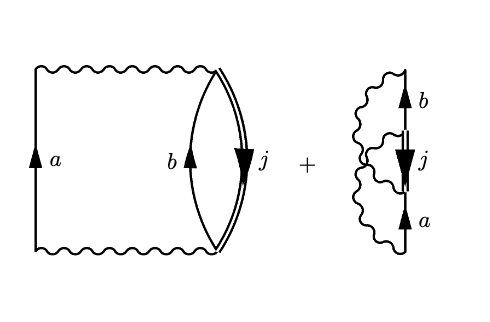} 
    \caption[Ground State Energy of Six-Site Model]{Interacting second-order Goldstone self-energy diagrams (iQPMP2 self-energy). Time is defined upwards, particle states are labelled $a, b$ and hole states are labelled $j$.}
    \label{fig:sc-se}
\end{figure}
Updating the second-order MP self-energy to account for the quasi-particle renormalization gives the expression for the interacting QPMP2 self-energy
\begin{equation}
    \Sigma^{i\QPMP2}_{ii}(\omega) = \frac{1}{2}\sum_{abj}\frac{|\braket{ij||ab}|^2}{\omega+\epsilon^{\QP}_{j}-\epsilon_{a}-\epsilon_{b}} \ .
\end{equation}
This renormalization is represented in Figure \ref{fig:sc-se}, where the hole lines are now dressed. By the same procedure, we may take the trace of the updated self-energy evaluated at the quasi-particle energy to give the interacting QPMP2 (iQPMP2) correlation energy as 
\begin{equation}\label{eq:sc-qpmp2}
    E^{i\QPMP2}_c = \frac{1}{4}\sum_{ijab} \frac{|\braket{ij||ab}|^2}{\epsilon^{\QP}_i+\epsilon^{\QP}_{j}-\epsilon_{a}-\epsilon_{b}} \ .
\end{equation}
This represents no increase in scaling as the quasi-particle energies are already found by solution of the quasi-particle equation (\ref{eq:qp_eqn}). As will be shown, this expression substantially improves the dissociation curves obtained for molecular systems, but reduces the performance of QPMP2 for Hubbard models.

Many-Body Brillouin-Wigner perturbation theory is constructed by defining the following projector 
\begin{equation}
    R_{0} = \sum_{m\neq 0} \frac{\ket{\Phi_m}\bra{\Phi_m}}{E_0-E^{(0)}_{m}} ,
\end{equation}
where $E_0$ is the exact ground state energy. The perturbative expression for the ground state energy within this formalism is given by 
\begin{equation}
    E_0 = E^{\HF}_{0} + \braket{\Phi_0|V\sum^{\infty}_{n=1}(R_0V)^n|\Phi_0} .
\end{equation}
Expanding self-consistently to second-order for the correlation energy gives
\begin{gather}
\begin{split}
    E^{\BW2}_0 = E^{\HF}_0+\frac{1}{4}\sum_{abij} \frac{|\braket{ij||ab}|^2}{E^{\BW2}_0- E^{(0)}_0-\Delta^{ab}_{ij}} \ . 
\end{split}
\end{gather}
This is the many-body equivalent of the Dyson series, but it is not clear what the energy difference of the denominator should be in order to decouple these equations. Additionally, the expression is not size-extensive and is evidently problematic for extended systems.\cite{keller2022regularized} There have been a few attempts to introduce a size-extensive implementation of second-order BW perturbation theory via ad-hoc procedures of defining this energy difference as a correlation energy per electron. From the size-extensive, interacting quasi-particle MP2 correlation energy, we can view the combination of all the energy differences between the zeroth order energy and the ground state energy $E^{\BW2}_0-E^{(0)}_0$ as the self-energy of each occupied state
\begin{gather}
\begin{split}
    E_0 &= E^{\HF}_{0} + \frac{1}{4}\sum_{abij} \frac{|\braket{ij||ab}|^2}{\Sigma^{\QPMP2}_{ii}(\epsilon^{\QP}_i)+\Sigma^{\QPMP2}_{jj}(\epsilon^{\QP}_j)-\Delta^{ab}_{ij}}\\ .
\end{split}
\end{gather}
This is exactly the expression for the interacting quasi-particle MP2 correlation energy. Hence we can view QPMP2 as a size extensive Brillouin-Wigner perturbation theory. 

\section{Results For Hubbard Models}\label{hubbard}

The Hubbard Hamiltonian has been used for decades as a conceptually simple, yet insightful model to probe the nature strong electronic correlation and the metal-to-insulator transition.\cite{hubbard1964electron,essler2005one} The 1D periodic Hubbard model consists of a lattice of sites, each of which support a maximum of two electronic states (one spin up, one spin down). The model consists of a `hopping' term which couples nearest neighbour sites (favouring electron delocalization) and a repulsive interaction when two electrons are on the same site (favouring electron localization). The two terms provide an intuitive model of the behaviour of electrons in molecules and materials. In the site basis, the Hamiltonian is
\begin{equation}\label{eq:4.1}
    H=-t\sum_{ i\sigma}\left(c^{\dag}_{i\sigma}c_{i+1\sigma}+h.c.\right)+U\sum_{i}n_{i\uparrow}n_{i\downarrow} \ , 
\end{equation}
where imposing periodic boundary condition means $c^{\dag}_{N+1\sigma}=c^{\dag}_{1\sigma}$. The Hamiltonian can be parametrized by the ratio $\frac{U}{t}$, which provides a universal measure of the `correlation strength'. As $\frac{U}{t}$ increases, the on-site repulsion increases relative to the `hopping' term, favouring configurations where the electron density is localized about each site. By neglecting the correlated fluctuations in the electron density about each site, we obtain the HF Hamiltonian for the Hubbard model
\begin{equation}\label{eq:HF_ham}
    H^{MF} =-t\sum_{ i\sigma}\left(c^{\dag}_{i\sigma}c_{i+1\sigma}+h.c.\right) +\frac{U}{2}\hat{N}-\frac{U}{4}N_{s}
\end{equation}
where $\hat{N}=\sum_{i\sigma}n_{i\sigma}$ is the total number operator and $N_s$ is the number of sites. This will always be a linear function of $\frac{U}{t}$.

\subsection{The Hubbard Dimer}
The half-filled Hubbard dimer provides a simplified model of the hydrogen molecule in a minimal basis. It is an insightful system to understand the nature of strong correlation present in the chemical bond and there have been numerous studies of the Hubbard dimer as a benchmark for electronic correlation methods.\cite{marie2021perturbation,di2021scrutinizing,burton2021variations,romaniello2012beyond,romaniello2009self} The exact ground state energy of the Hubbard dimer is given by\cite{tomczak2007spectral,romaniello2009self} 
\begin{equation}
    E_0 = \frac{U}{2}- \frac{1}{2}\sqrt{U^2 + 16t^2} \ .
\end{equation}
To find the exact Green's function requires the diagonalization of the single and three particle Hamiltonians as well. Using these states, it is simple to construct the exact Green's function and polarization from their Lehmann representations. As the Hamiltonian is spin-independent, all expressions are spin-diagonal and the spin-index can be dropped.
\begin{figure*}
    \centering
    \subfloat[\centering Ground state energy of Hubbard dimer]{{\includegraphics[width=100mm, height = 75mm]{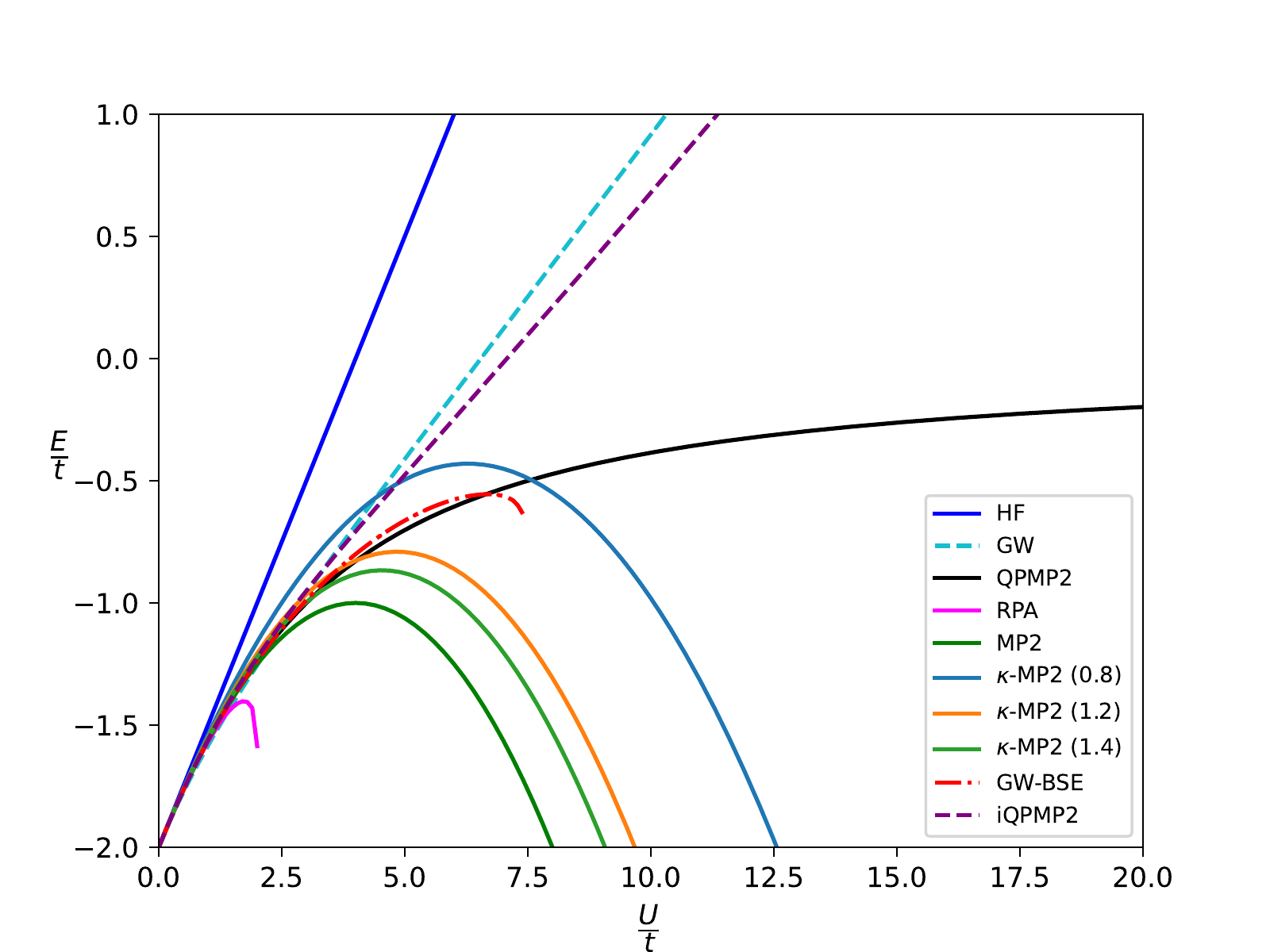}}}
    \hfill
    \subfloat[\centering Exact and $GW$ quasi-particle weights]{{\includegraphics[width=80mm, height = 60mm]{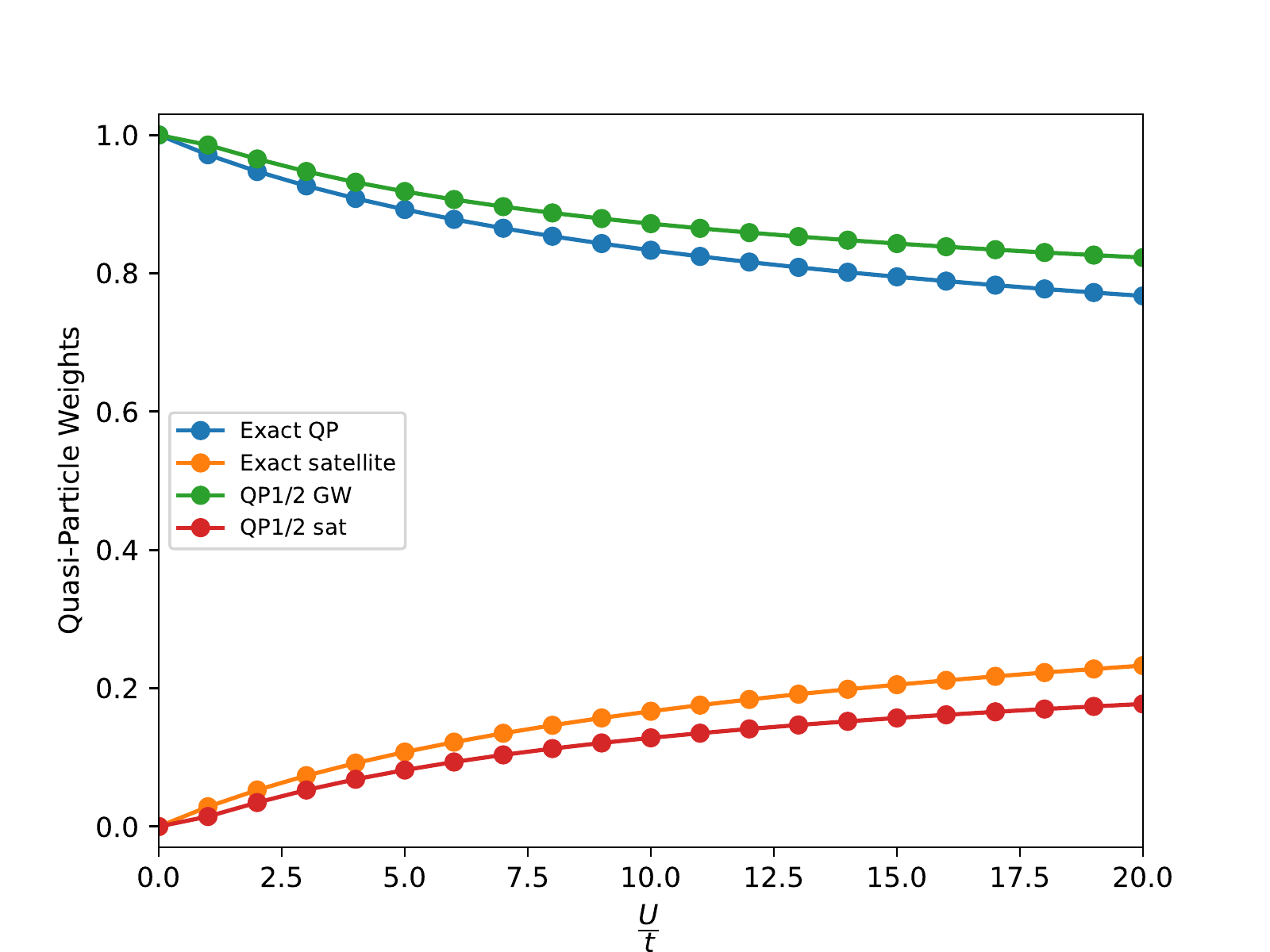} }}%
    \subfloat[\centering Exact and $GW$ quasi-particle energies]{{\includegraphics[width=80mm, height = 60mm]{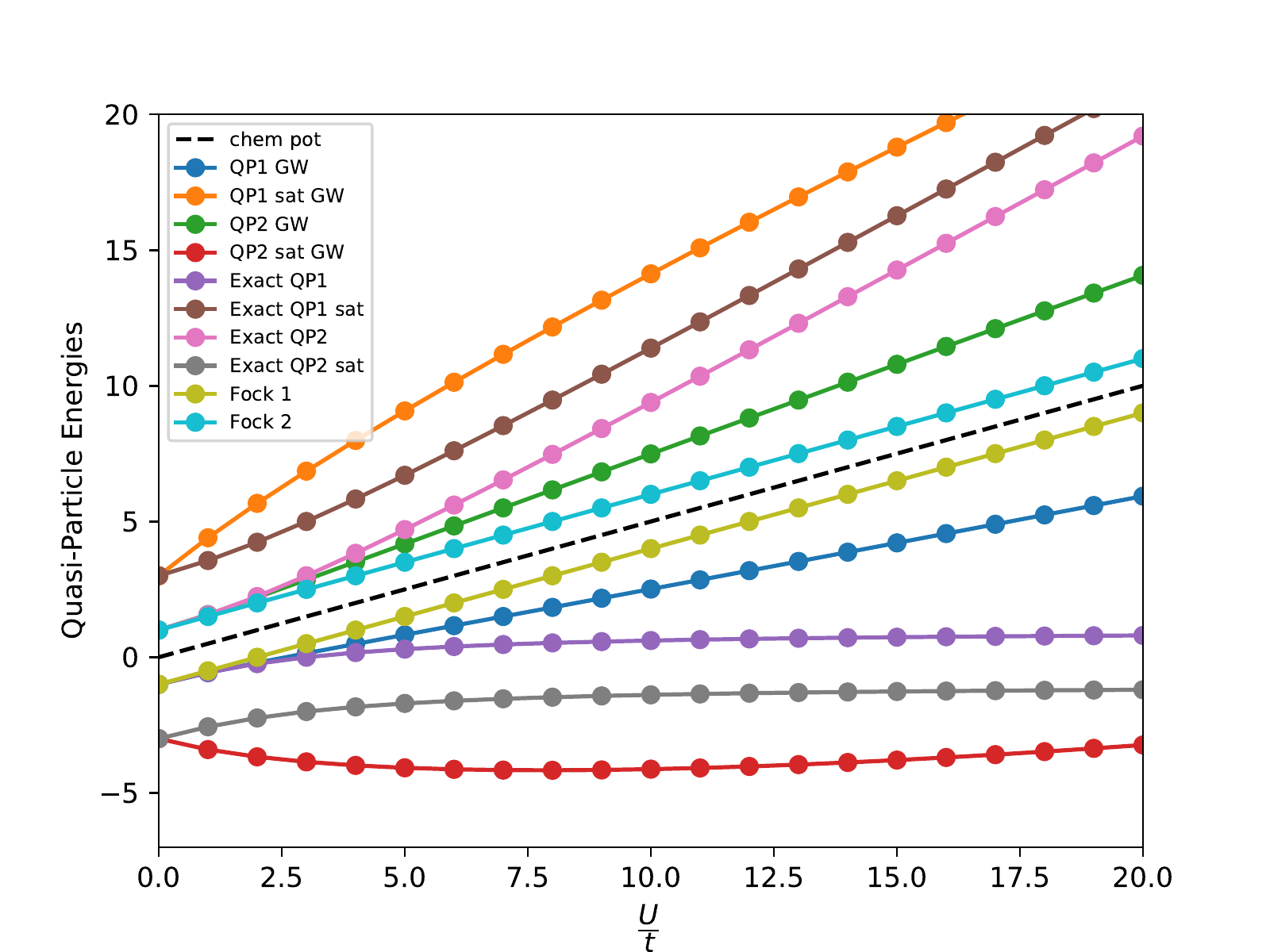} }}%
    \caption{(a) Ground state energy of the Hubbard dimer as a function of $\frac{U}{t}$. (b) Variation of the exact and $GW$ quasi-particle weights and (c) energies as a function of $\frac{U}{t}$ for the Hubbard dimer.}
    \label{fig:dimer}
\end{figure*}
From knowledge of the exact and HF Green's functions, we can invert Dyson's equation to find the exact exchange-correlation self-energy. In the diagonal plane-wave basis, the exact Feynman-Dyson self-energy is given by
\begin{gather}\label{eq:exact}
\begin{split}
    \Sigma_{11}(\omega)&= \frac{U^2}{4}\frac{1}{\omega-(\frac{U}{2}+3t)+i\eta}\\
    \Sigma_{22}(\omega)&= \frac{U^2}{4}\frac{1}{\omega-(\frac{U}{2}-3t)-i\eta} \ .
\end{split}
\end{gather}
The second-order Feynman-Dyson self-energy is given by 
\begin{gather}
    \begin{split}
        \Sigma^{(2)}_{pq}(\omega) = \frac{1}{2}\left(\sum_{abi} \frac{\braket{pj||ab}\braket{ab||qj}}{\omega+\epsilon_j-\epsilon_a-\epsilon_b+i\eta} \right .\\
        \left. + \sum_{ija} \frac{\braket{pa||ij}\braket{ij||qa}}{\omega+\epsilon_a-\epsilon_i-\epsilon_j-i\eta} \right) .
    \end{split}
\end{gather}
Clearly, we see that the second-order Feynman-Dyson self-energy is exact for the Hubbard dimer as $\braket{11|22} = \frac{U}{2}$ and $\epsilon_{1/2}= \frac{U}{2}\mp t$. The resulting Dyson equation returns the exact Green's function and therefore ground state energy and spectral function. Essentially all the dynamical and stationary state observables for the Hubbard dimer are reproduced exactly by the second-order approximation for the self-energy. Our expression for the second-order Goldstone self-energy for the Hubbard dimer gives
\begin{equation}
    \Sigma^{(2)}_{11}(\omega) = \frac{U^2}{4}\frac{1}{\omega-\left(\frac{U}{t}-3t\right)} \ .
\end{equation}

Solving the quasi-particle equations for this self-energy and taking the solutions with the largest weight gives the QPMP2 quasi-particle energies
\begin{gather}
    \begin{split}
        \epsilon^{\QP}_{1}= \epsilon^{\HF}_{1} +  \Sigma^{(2)}_{11}(\epsilon^{\QP}_1) = \frac{U}{2}+t-\frac{1}{2}\sqrt{U^2+16t^2}\\
        \epsilon^{\QP}_{2}= \epsilon^{\HF}_{2} +  \Sigma^{(2)}_{22}(\epsilon^{\QP}_2) = \frac{U}{2}-t+\frac{1}{2}\sqrt{U^2+16t^2}
    \end{split}
\end{gather}

\noindent The corresponding QPMP2 ground state energy is given by  
\begin{gather}
    \begin{split}
        E^{\QPMP2}_0 &= \frac{U}{2}-2t +\Sigma^{(2)}_{11}(\epsilon^{\QP}_1) \\
        &= \frac{U}{2} -2t - \frac{U^2}{4}\left[\frac{1}{2t+\frac{1}{2}\sqrt{U^2+16t^2}}\right] \ ,
    \end{split}
\end{gather}
which is also exact for the Hubbard dimer as
\begin{equation}
    2t + \frac{U^2}{4}\left[\frac{1}{2t+\frac{1}{2}\sqrt{U^2+16t^2}}\right] \equiv \frac{1}{2}\sqrt{U^2+16t^2} \ .
\end{equation}
Therefore, QPMP2 also reproduces the exact ground state energy of the Hubbard dimer. 
The MP2 correlation energy can be found by simply evaluating the second-order Goldstone self-energy at the Fock single-particle energy and diverges due to the multireference nature of the strongly correlated solution.\cite{phillips2014communication} The CCSD equations can be solved analytically, reproducing the exact ground state wavefunction and energy over the entire correlation regime as it is exact for two electron systems. 

Figure \ref{fig:dimer}(a) depicts the ground state energy of the Hubbard dimer as a function of $\frac{U}{t}$ for a variety of different electronic structure methods. We compare QPMP2 to regular GF2, RPA, GW, GW-BSE, CCSD, MP2 and HF theories. These results provide a qualitative insight into the relative behaviour of these methods when accounting for bond dissociation processes. 

The RPA energy provides a slight improvement over the HF approximation for values of $\frac{U}{t}< 1.8$, but the presence of complex excitation energies for $\frac{U}{t} \geq 2$ results in a discontinuous curve.\cite{scuseria2008ground,di2021scrutinizing} The sudden onset of complex excitation energies as the correlation strength increases can be related to the change in the character of the HF Hessian matrix.\cite{bulik2015can} The RPA matrix is the HF Hessian matrix and if this matrix is positive-definite, then all the excitation energy eigenvalues will be real.  Therefore, the presence of complex excitations is a consequence of the fact that the HF solution is no longer at a stable minimum due to the multireference character of the correlated ground state and is related to the singlet-triplet instability.\cite{yamada2015singlet} The divergent behaviour of RPA when accounting for ground state electronic correlation is a consequence of the fact that artificial boson commutation relations are imposed on the fermion electron-hole pair operators.

The GW approximation sees a significant improvement over HF, by correlating single-particle states via the `first-order' renormalized self-energy of Hedin.\cite{hedin1965new,caruso2013self,rostgaard2010fully} The GW energy is reasonable within the weak correlation regime (up to values of $\frac{U}{t} \leq 2.6$) but fails in the strong-correlation limit, displaying an approximately linear correlation energy.\cite{di2021scrutinizing} 

Figure \ref{fig:dimer}(b) displays how the weight of the quasi-particle and satellite solutions vary as the correlation strength increases. It is clear that the weight of the quasi-particle state is dominant over the entire regime for both GW and the exact system. Figure \ref{fig:dimer}(c) shows the variation in the quasi-particle weights with the correlation strength. From this plot we see that the band gap increases as $\frac{U}{t}$ increases for both QPMP2 and GW. The `exact' band-gaps change much more significantly as a function of $\frac{U}{t}$ when compared to both the GW and HF approximations. We also see that the variation in the energy of the satellite solutions across the entire correlation regime is much less significant than for that of the quasi-particle solutions. The entire spectrum of single-particle states is symmetric about the chemical potential due to particle-hole symmetry.

\begin{figure*}
    \centering
    \subfloat[\centering Ground state energy of periodic six-site Hubbard model]{{\includegraphics[width=100mm, height = 75mm]{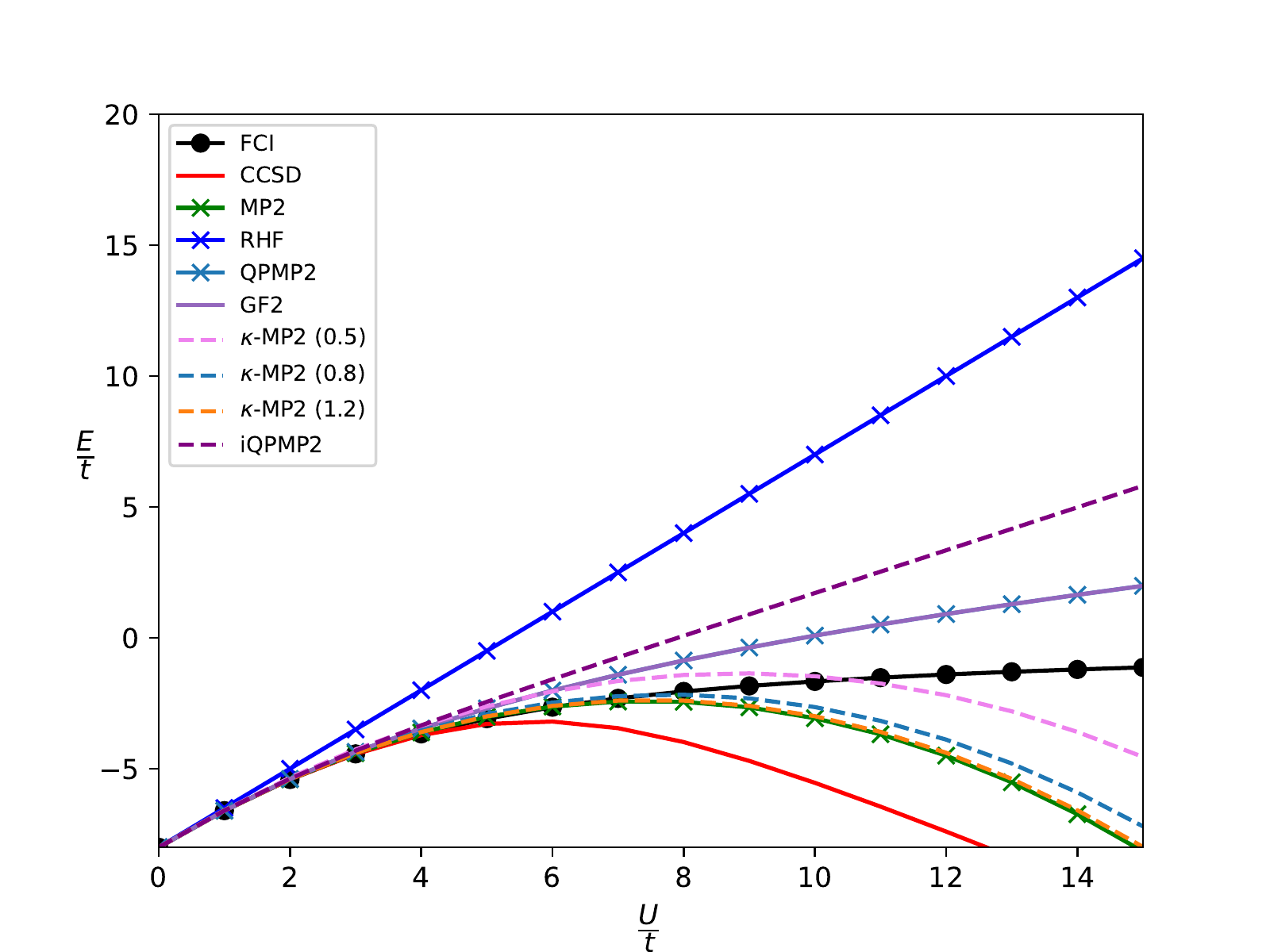}}}%
    \hfill
    \subfloat[\centering GF2 quasi-particle weights]{{\includegraphics[width=80mm, height = 60mm]{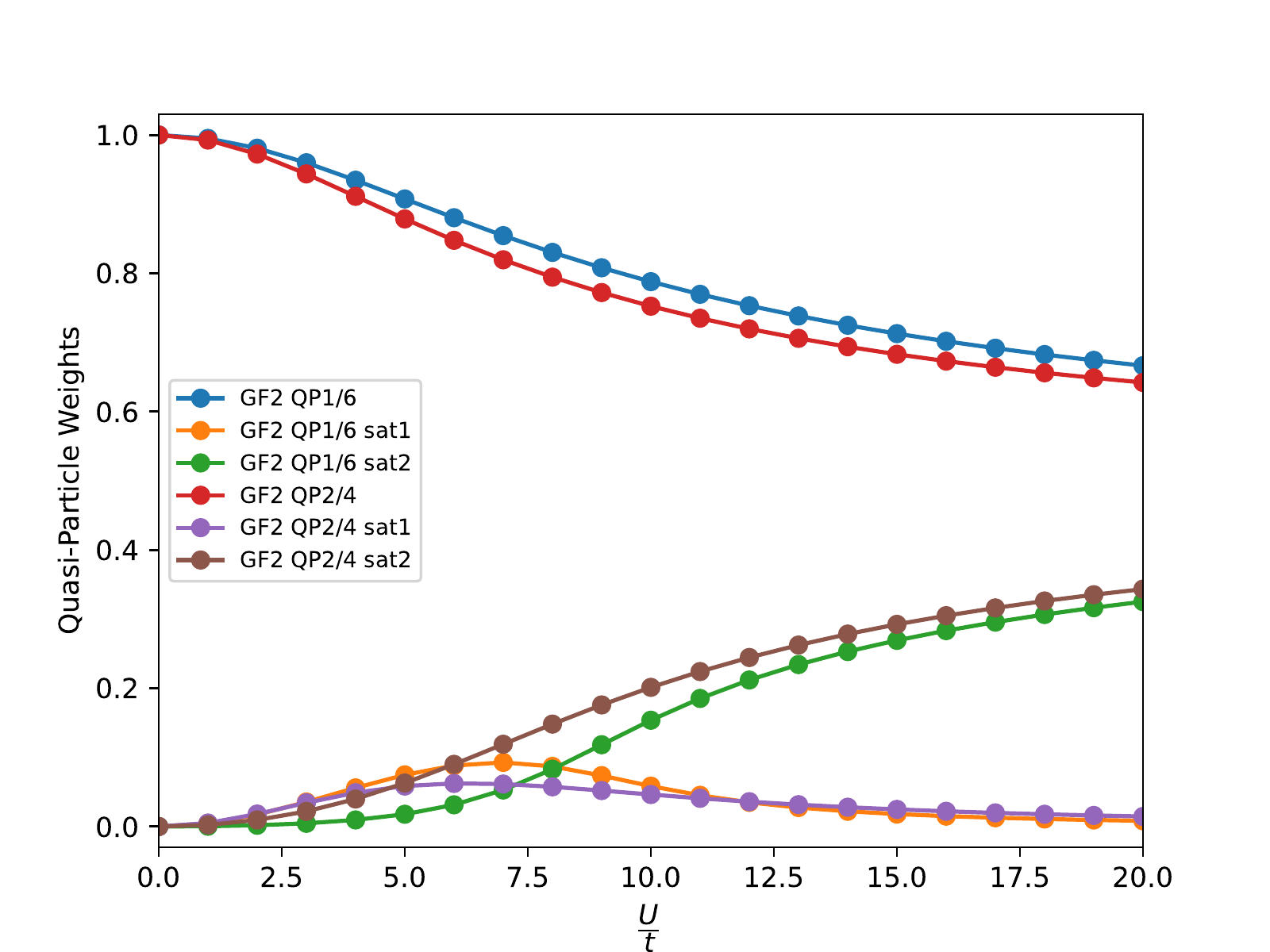}}}%
    \subfloat[\centering GF2 quasi-particle energies]{{\includegraphics[width=80mm, height = 60mm]{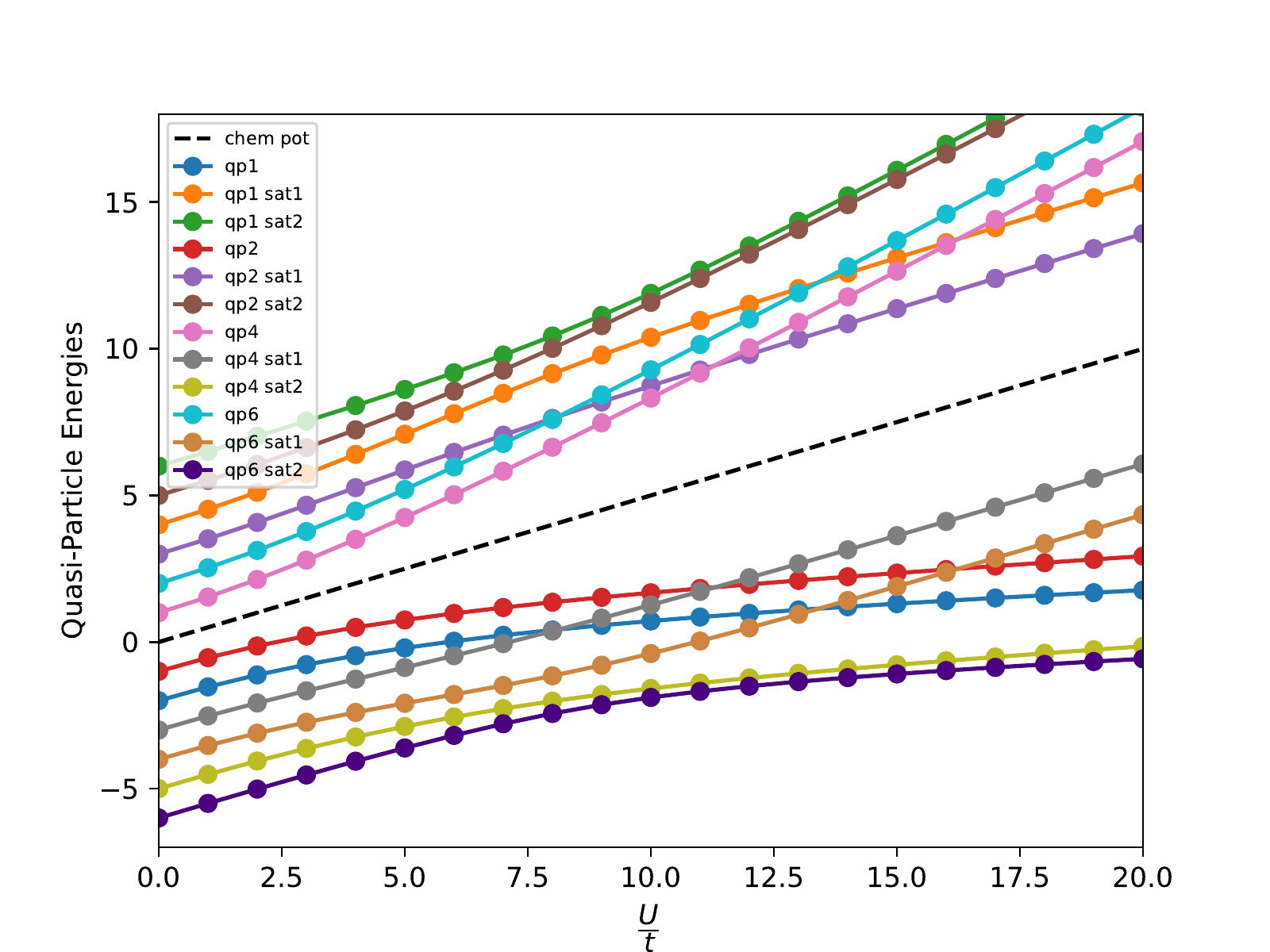}}}%
    \caption{(a) Ground state energy of the periodic six-site Hubbard model as a function of $\frac{U}{t}$. (b) Variation of the GF2 quasi-particle weights and  (c) energies as a function of $\frac{U}{t}$ for the periodic six-site Hubbard model.}
    \label{fig:6site}
\end{figure*}

Similar to the relationship between RPA and HF, the GW-BSE method is dependent on the nature of the GW solution. It shows a marked improvement over the GW approximation, matching the exact energy over the region where GW performs well, and is accurate up to $\frac{U}{t}<7$. However, in the strong correlation regime, the GW-BSE excitation energies become complex as the GW solution becomes unstable. The RPA and GW-BSE approaches give complex excitation energies in the strong correlation limit where the HF/GW starting point is no longer at a stable minimum.\cite{li2020ground,maggio2016correlation} Therefore, we suspect that the GW-BSE method applied to real molecular systems will likely exhibit similar divergences to those of RPA for the ground state energy.

The GF2 and QPMP2 approximations are exact for the Hubbard dimer. The second-order approximation corresponds to an infinite order resummation of the MP2 diagrams via the Dyson equation. The fact that QPMP2 also reproduces the exact ground state energy is even more intriguing as it suggests that the Hubbard dimer quasi-particle solutions can be considered as `exact' single-particle states, and that the satellites do not contribute to the energy. Interestingly, the interacting QPMP2 expression is no longer exact for the Hubbard dimer system and substantially decreases the accuracy of the approach, resulting in a correlation energy that is similar to that obtained from the GW approximation.

The $\kappa$-regularization scheme also diverges akin to MP2 for the Hubbard dimer. Figure \ref{fig:dimer}(a) displays the divergence of $\kappa$-MP2 for a range of $\kappa$ values from $0.8$ to $1.4$. From equation \ref{eq:kappa}, we see that as $\kappa$ tends to zero we recover the HF energy as as $\kappa\to\infty$ we obtain the MP2 correlation energy. We see that no typical value of $\kappa$ will regularize MP2 for the Hubbard dimer and all regularization schemes based on this approach fail for larger Hubbard models (Section \ref{sub:hub}). This is because the divergence of MP2 for the Hubbard model is actually due to the numerator rather than the denominator as $\Delta^{ab}_{ij}$ is constant over the entire range of $\frac{U}{t}$. Similar results are observed for the second-order perturbative DSRG expression of equation \ref{eq:DSRG2}. Therefore, one must be careful when using Hubbard models to understand electronic correlation in molecules and materials.

\subsection{Periodic Six-Site Hubbard Model}\label{sub:hub}

The half-filled periodic six-site Hubbard model is a closed shell system where a single HF reference provides a reasonable starting point. Figure \ref{fig:6site}(a) shows the plot of the ground state energy of the six-site model as a function of the correlation strength for a number of different approximate methods. The FCI and CCSD calculations were performed using the PySCF module.\cite{sun2018pyscf}

 As is the case for the dimer, the HF solution diverges from the exact solution for $\frac{U}{t}\geq 1$, where the electronic motion is no longer weakly correlated, and drastically fails at larger correlation strength. Of the traditional post HF electronic structure methods widely used, it would appear that MP2 performs best. However, the MP2 energy diverges, due to the highly multireference nature of the FCI wavefunction and begins to diverge for $\frac{U}{t}\ge4$. This is verified in Figure \ref{fig:6site}(b), as the quasi-particle weights from the GF2 calculation also begin to change significantly around $\frac{U}{t}\approx4$. Again, $\kappa$-MP2 also diverges for the six-site model and does not provide a regularization of MP2. The curves show that ultimately the $\kappa$-MP2 regularization scheme provides little to no improvement over MP2 for the Hubbard model. Again, this divergence is due to the numerator rather than the denominator of the MP2 expression. Hence, the Hubbard model represents a significantly challenging condensed matter physics system for benchmarking regularization schemes and traditional electronic structure theories.\cite{burton2022exact}

 CCSD performs well in the weak correlation regime for values of $\frac{U}{t} \leq 4$. However, it also diverges in the intermediate-strong correlation limit as the weight of higher excitations in the FCI begins to dominate the doubles amplitude. This is a spectacular failure as CCSD does not provide a qualitatively accurate description of the metal-to-insulator transition present in the FCI.\cite{phillips2014communication,scuseria2008ground,evangelista2018perspective}

The GF2 and QPMP2 methods give the same ground state energy for the six-site model across the entire correlation regime. Both perform similarly to CCSD and MP2 in the weak correlation regime as all methods begin to move away from the FCI energy at $\frac{U}{t}\approx 4$. However, the ultimate difference is that neither the GF2 or QPMP2 energy diverges in the strongly correlated regime and therefore, they provide a qualitatively accurate description of the strongly correlated system. As a result, the metal-to-insulator transition is captured and the divergences of MP2 are cured. Again, it is noted that the interacting QPMP2 correlation energy expression (\ref{eq:sc-qpmp2}) decreases the accuracy of the QPMP2 expression for the six-site Hubbard model.

\begin{figure*}[ht]
    \centering
    \subfloat[\centering H$_2$ molecule (cc-pVDZ)]{{\includegraphics[width= 85mm, height = 67.5mm]{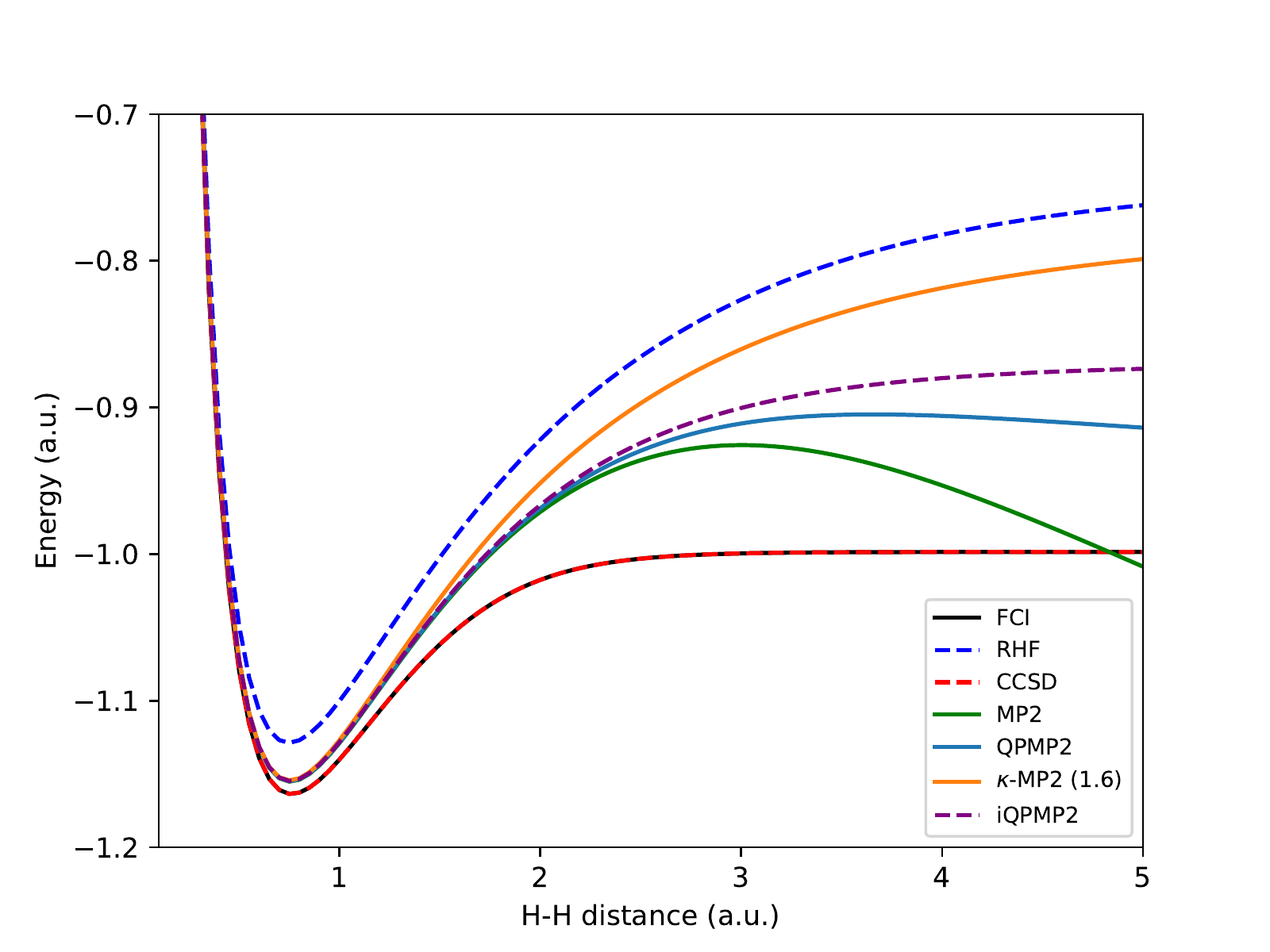}}}%
    \hfill
    \subfloat[\centering N$_2$ molecule (STO-3G)]{{\includegraphics[width= 85mm, height = 67.5mm]{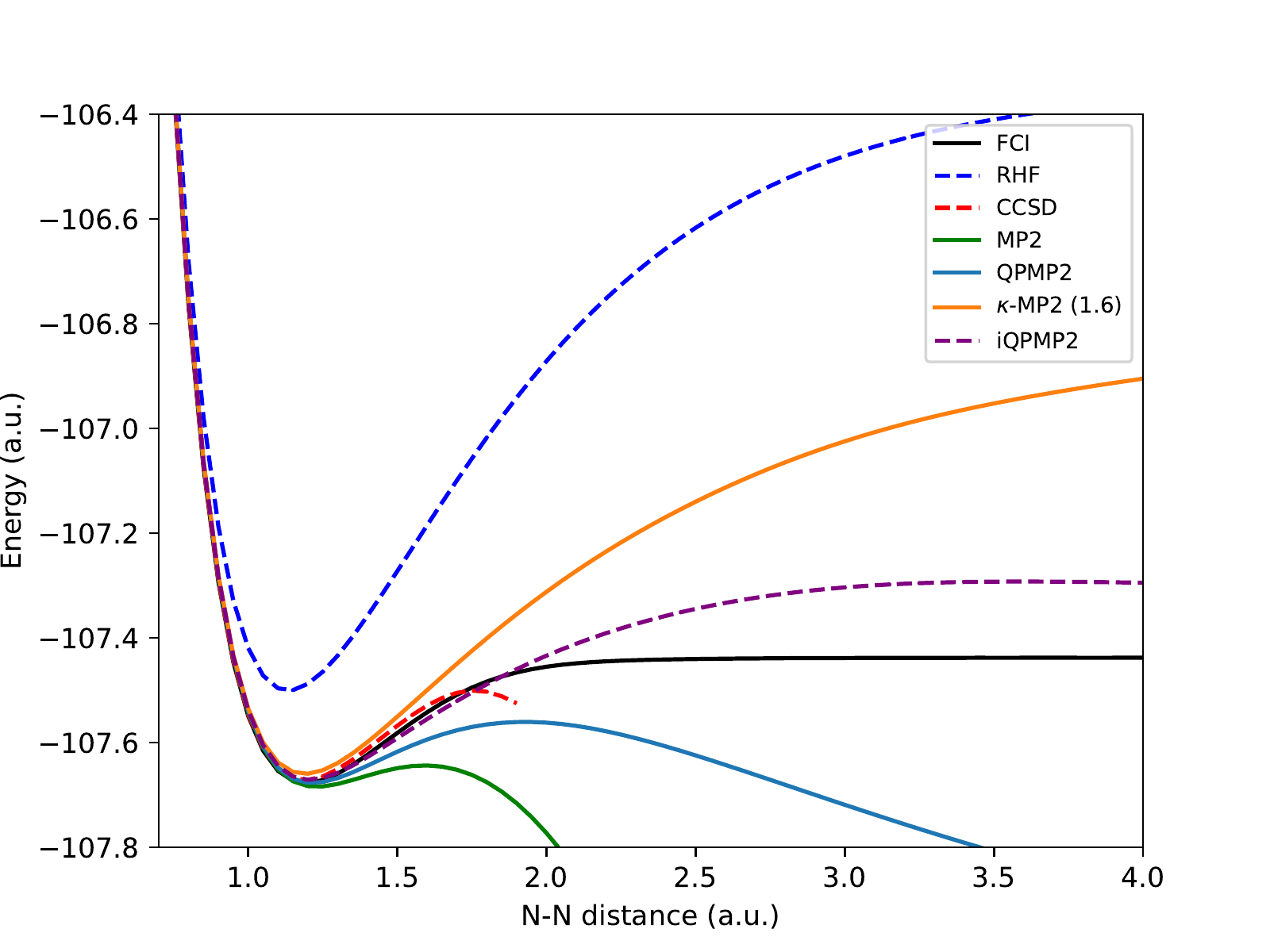}}}%
    \hfill
    \subfloat[\centering Linear H$_6$ (STO-3G)]{{\includegraphics[width = 85mm, height = 67.5mm]{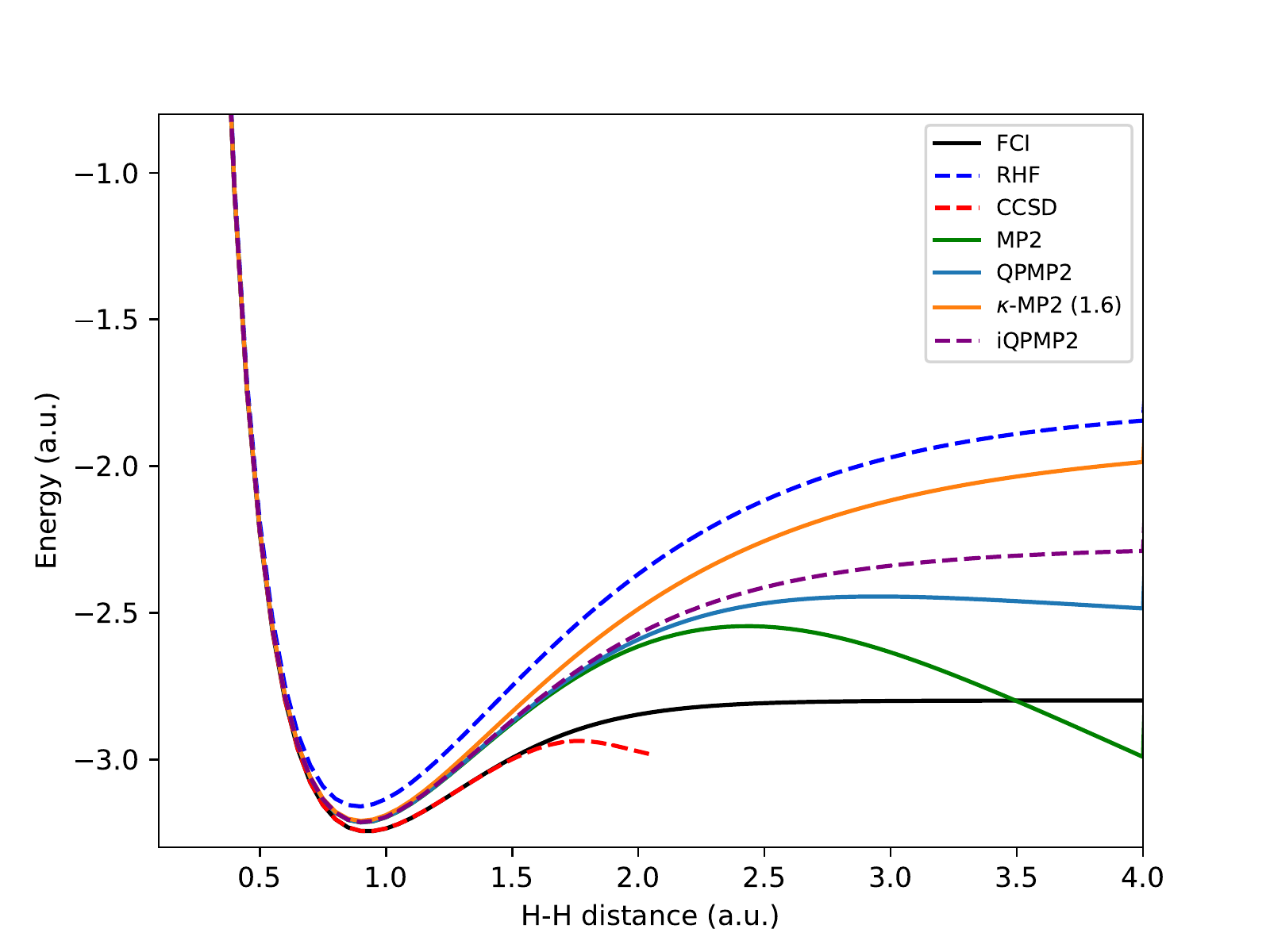}}}%
    \hfill
    \subfloat[\centering Ring H$_6$ (cc-pVDZ)]{{\includegraphics[width = 85mm, height = 67.5mm]{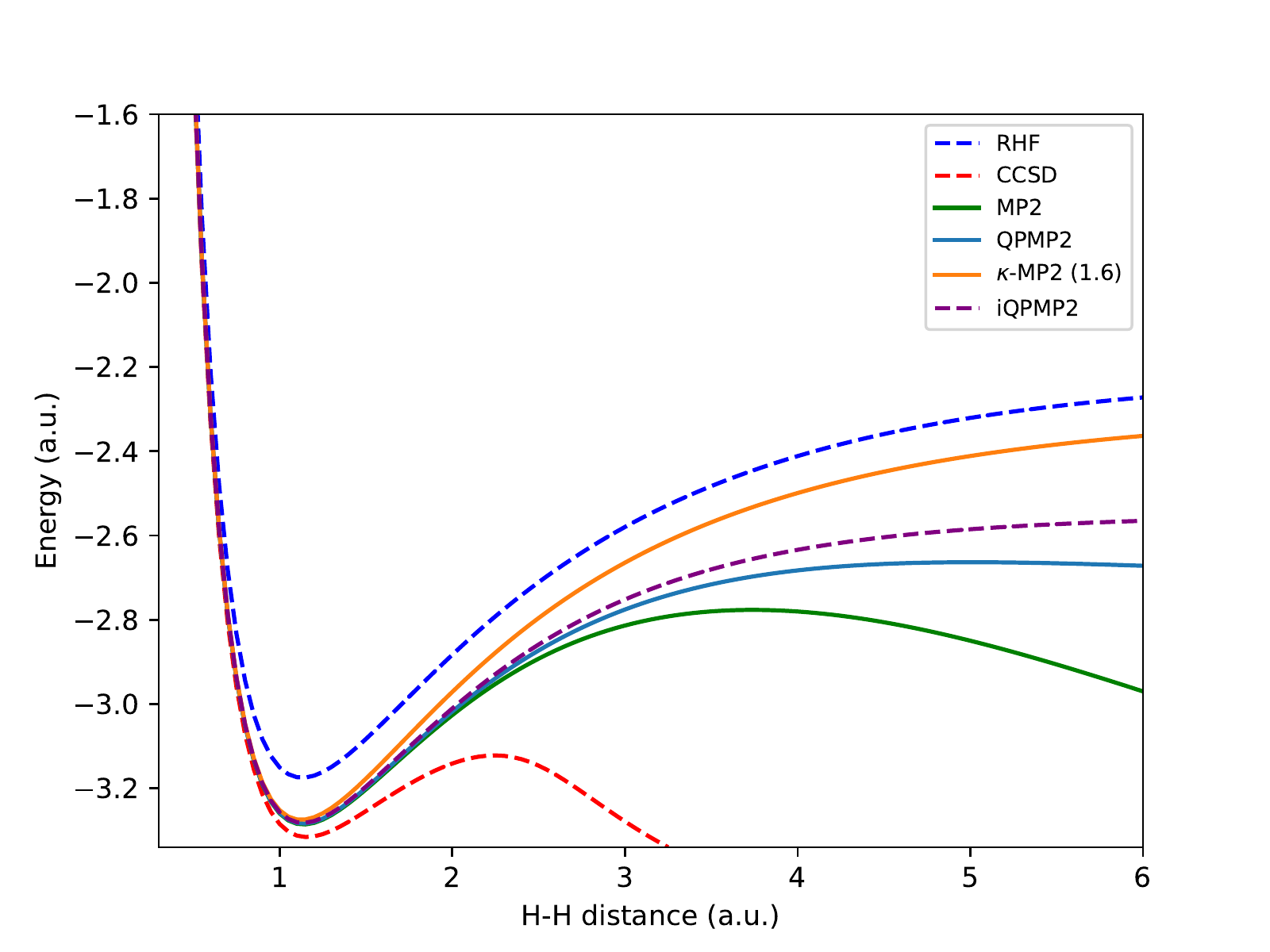}}}%
    \caption{Ground state energy of H$_2$ molecule in STO-3G basis (a), N$_2$  molecule in STO-3G basis (b), linear H$_6$ in STO-3G basis (c) and H$_6$ ring in cc-pVDZ basis (d).}
    \label{fig:h6}
\end{figure*}

Figure \ref{fig:6site}(b) contains the variation of the GF2 quasi-particle weights as a function of the correlation strength. Through the variations in the quasi-particle weights as a function of correlation strength, we can understand at the nature of the different quasi-particle solutions and the QPMP2 correlation energy. As can be seen, the weights of the satellite solutions begin to become increasingly important for $\frac{U}{t}\geq 8$. This is the hallmark of strong correlation, where many-body effects are increasingly present. However, the weight of the first satellite solutions is effectively zero across the entire correlation regime and can be effectively ignored altogether. The quasi-particle energies in Figure \ref{fig:6site}(c) are again symmetric about the chemical potential as the self-energy preserves the particle-hole symmetry. Their variation with the correlation strength again highlights the increase in the band gap as $\frac{U}{t}$ increases. It is the correlation embedded in these quasi-particle states which leads to the QPMP2 regularization of MP2 for the Hubbard model.

Analogous behaviour is observed for the linear eight- and ten-site Hubbard models which also display the characteristic divergences of CCSD, MP2, $\kappa$-MP2 and DSRG-PT2. Again, QPMP2 theory qualitatively reproduces the metal-to-insulator transition present in the FCI solution and is non-divergent over the entire range of $\frac{U}{t}$. The performance of QPMP2 for the Hubbard dimer and larger Hubbard models indicate that this regularization is sufficient to cure the divergences of MP2 within the strongly correlated regime, providing a qualitatively accurate description of the metal-to-insulator transition.

\section{Results for Molecular Systems}\label{molecules}

In this section, we apply QPMP2 to a series of molecular systems which demonstrate the transition from weak to strong correlation present in quantum chemistry. All dissociation curves level off at large distances unless stated otherwise. Figure \ref{fig:h6}(a) shows the ground state potential energy surface for H$_2$ in the cc-pVDZ basis set. Clearly CCSD is exact for any two-electron system and recovers the FCI dissociation curve and exact ground state wavefunction.


The well-known divergence of MP2 for the simplest molecular system is a result of energy denominator $\Delta^{ab}_{ij}$ tending to zero as the distance between the hydrogen nuclei increases. The dissociation curve for QPMP2 cures the divergence of the MP2 solution whilst showing an identical energy surface around the equilibrium configuration. The single-shot calculation is enough to open up the band gap and regularize the MP2 energy in the strongly correlated regime.  

However, the method substantially underestimates the correlation energy in the dissociation limit for H$_2$ as it simply regularizes the MP2 correlation energy. This is in stark contrast to the case for the Hubbard dimer, where both QPMP2 and GF2 are exact over the entire range of $\frac{U}{t}$. This again reiterates that using the Hubbard dimer as a simple benchmark system for electronic structure methods may not always be appropriate. The non-zero slope at dissociation of the QPMP2 dissociation curve (which levels off at large bond distances) is completely removed in the interacting QPMP2 correlation energy expression. The interacting QPMP2 energy introdcues no additional scaling as all the quasi-particle energies have already been calculated from the quasi-particle equation and is as a result of `iterating' the self-energy once to include the effects of the renormalized hole states (Section \ref{BW}). It has been shown in Section \ref{hubbard} that the interacting QPMP2 approach for Hubbard model systems reduces the performance of the QPMP2 correlation energy. However, we find this update of the self-energy to be necessary to correctly describe bond dissociation processes. The observed improvement resulting from iteration of the self-energy should be noted as a fundamental difference between the electronic correlation in real molecules and Hubbard models. We would like to comment on the similarity between our interacting QPMP2 expression (\ref{eq:sc-qpmp2}) and the expression obtained in the recent work of Carter-Fenk and Head-Gordon on a size consistent Brillouin-Wigner correlation energy expression involving dressed occupied orbital eigenvalues.\cite{carter2023tensor}

In addition, the $\kappa$-MP2 ground state energy of H$_2$ is shown for a typical value of $\kappa = 1.6$ reported in the literature.\cite{loipersberger2021exploring, shee2021regularized} As we can see QPMP2, $\kappa$-MP2 and MP2 are essentially identical near the equilibrium nuclei configuration. However, as the correlation strength increases towards dissociation, both QPMP2 methods recover a much larger correlation energy. The divergences of $\kappa$-MP2 seen for the Hubbard models do not appear for H$_2$ as the divergence of MP2 in this case is due to the denominator $\Delta^{ab}_{ij}$ rather than the numerator. 

We also benchmark our method on characteristic systems for which CCSD and CCSD(T) diverge. The linear H$_6$ and H$_{10}$ chains as well as the cyclic H$_6$ ring are such systems where effects of strong correlation present difficulties. The spectacular failures of CCSD and CCSD(T) for these systems is concerning and highlights the current challenges strong correlation presents for scalable electronic structure theories. 

For the linear H$_6$ system (Figure \ref{fig:h6}(c)) we see the same divergent behaviour of MP2 due to the vanishing band gap that is renormalized by QPMP2 theory. However, again the correlation energy is overestimated at `dissociation' and QPMP2 simply acts as a regularization of MP2. The interacting QPMP2 energy removes the non-zero slope of the QPMP2 energy at dissociation. Again, $\kappa$-MP2 also regularizes MP2 in contrast with the Hubbard model results of Section \ref{hubbard}, but does not recover the correlation energy is the strongly correlated regime. CCSD diverges for both H$_6$ systems and linear H$_{10}$ due to the onset of strong correlation as the H-H distance increases.  

Figure \ref{fig:h6}(d) shows the ground state energy of the H$_6$ ring as a function of the breathing mode within the cc-pVDZ basis. We see similar quantitative results as for linear H$_6$ with the regularization scheme again providing a qualitatively accurate energy at `dissociation'. However, again we see no improvement over MP2 about the equilibrium configuration as the curves are essentially identical. 
 Results for linear H$_{10}$ (not shown here) also display the same behaviour as the linear and cyclic H$_6$ systems. We note that the quantitative accuracy of CCSD for these systems is unrivalled within the weak-intermediate correlation regime. 
 
 Similar qualitative results for H$_6$, H$_{12}$ and H$_{32}$ lattices have been demonstrated by the GF2 method of Zigid et al.\cite{kananenka2016efficient,phillips2014communication,pokhilko2021evaluation,neuhauser2017stochastic} However, their implementation is based on the finite temperature formalism, where the Green's function and self-energy are constructed in the imaginary frequency and imaginary time domains.
At finite temperature, the Green's function is anti-periodic in imaginary time with a period of the inverse temperature $\beta$ and can be expanded as a Fourier series over discrete Matsubara frequencies. As a result, the ground state energy is expressed as a Matsubara frequency summation and their method utilises several properties of the Green's function which are not present at zero temperature.\cite{Quantum} Additionally, their implementation is based on the renormalized Feynman-Dyson perturbation theory to perform self-consistent calculations which are much more computationally involved.

Finally, we discuss our regularization for the ground state energy of N$_2$ which presents a significant challenge at dissociation due to the highly correlated nature of the triple bond. In Figure \ref{fig:h6}(b), we see that CCSD performs well until it diverges in the dissociation limit. MP2 theory fails completely for the N$_2$ molecule as it barely reproduces the minimum in the energy surface before it diverges. In contrast, $\kappa$-MP2 provides an efficient regularization of MP2, but is less accurate near the equilibrium configuration. 

For N$_2$, we see the biggest difference between the interacting and single-shot QPMP2 correlation energy. Single-shot QPMP2 substantially improves the qualitative performance of MP2, clearly displaying a minimum. However, it also diverges beyond 2 a.u. as the nature of the electronic correlation exhibited by N$_2$ near dissociation is not described within this scheme. In stark contrast, iQMP2 completely cures the divergences of second-order perturbation theory for N$_2$ and provides an accurate potential energy surface which dissociates to the correct limit. From Figure \ref{fig:h6}(b), the dissociation curve of iQMP2 matches the FCI solution at around 1.9 a.u. and outperforms CCSD within the weak-intermediate correlation regime whilst remaining non-divergent at dissociation. However, the potential energy of iQMP2 at dissociation is again overestimated. This is a spectacular result that demonstrates the power of our second-order regularization scheme for molecular systems.

\section{Conclusions and Outlook}\label{conclusion}
We have introduced a scalable, size-consistent regularized second-order correlation method based on the single-particle Green's function which can be cast as a size-extensive Brillouin-Wigner perturbation theory. We have also outlined its connections to different regularized expressions that have been explored in the literature, with the additional advantage that QPMP2 represents a physically meaningful regularization scheme. We have shown that QPMP2 is capable of curing the divergences of MP2 in the strongly correlated regime for a wide range of cases and that the method is capable of qualitatively describing the metal-to-insulator transition. However, gaining a quantitatively accurate description of strong correlation within the Green's function formalism is not straightforward and requires the rigorous inclusion of three-body interactions which will likely involve higher order Green's functions. Future work must be carried out to understand whether the quantitative accuracy required for quantum chemistry can be achieved within the Green's function formalism. There is currently no Green's function based approach in the literature that can compete with the quantitative accuracy of CCSD or CCSD(T) for ground state energies and dissociation curves. A major open question is whether the divergences of CCSD(T) can be cured by exploring the connections between the Green's function and Equation-of-Motion Coupled Cluster theory.\cite{tolle2022exact}




\appendix

\nocite{*}

\section*{Author Information}

\noindent\textbf{Corresponding Author:} David P. Tew
\\
\textbf{Email:} \url{david.tew@chem.ox.ac.uk}
\\
\textbf{ORCID:} \url{https://orcid.org/0000-0002-3220-4177}
\\
\textbf{Notes:}
The authors declare no competing financial interest.
\vspace{30mm}
\section*{References}
\bibliography{qpmp2}

\begin{figure*}[ht]
    \centering
    \includegraphics{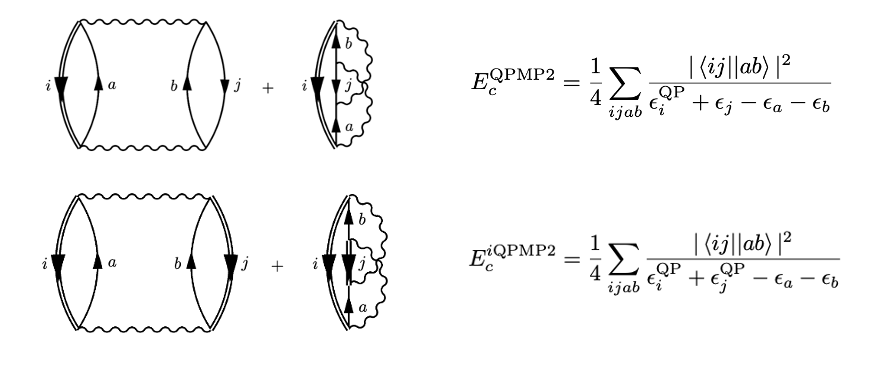}
\end{figure*}

\end{document}